\documentclass[traditabstract]{aa}
\usepackage[]{srcltx}
\usepackage{txfonts}
\usepackage{soul}
\usepackage{amssymb}
\usepackage[scriptsize,bf]{caption}
\usepackage{float}
\usepackage{placeins}
\usepackage{xspace}
\usepackage{natbib}
\usepackage{textgreek}
\usepackage{multirow}
\usepackage{rotating}

\newcommand{\rsun}{\mathrm{R}_{\mathrm{o}}}
\newcommand{\vsun}{\mathrm{\Omega}_{\mathrm{o}}}
\newcommand{\km}{\,\mathrm{km}}
\newcommand{\kms}{\,\frac{\km}{\mathrm{s}}}
\newcommand{\jgrad}{\,\frac{\km}{\mathrm{s}\cdot\mathrm{kpc}}}
\newcommand{\pc}{\,\mathrm{pc}}
\newcommand{\kpc}{\,\mathrm{kpc}}
\newcommand{\msun}{\mathrm{M}_{\odot}}

\newcommand{\mw}{\textsc{MW}\xspace}

\newcommand{\abs}[1]{\left|{#1}\right|}

\newcommand{\br}[1]{\left({#1}\right)}

\newcommand{\ud}[1]{\mathrm{d}{#1}}
\renewcommand{\exp}[1]{\mathrm{e}^{#1}}
\renewcommand{\sq}[1]{\left[{#1}\right]}
\newcommand{\Q}[1]{``#1''}
\newcommand{\mexhat}{\Q{mexican hat}\xspace}
\newcommand{\crv}[1]{\textsc{[#1]}\xspace}
\usepackage[usenames,dvipsnames]{color}
\usepackage[normalem]{ulem}
\newcommand{\xxout}[1]{}
\newcommand{\rem}[1]{}
\newcommand{\remv}[1]{}
\newcommand{\new}[1]{#1}
\newcommand{\add}[1]{#1}
\newcommand{\refrm}[1]{#1}

\newcommand{\secref}[1]{Sect.\,\ref{#1}}

\newcommand{\figref}[1]{Fig.\,\ref{#1}}
\newcommand{\eqref}[1]{Eq.\,\ref{#1}}
\newcommand{\newII}[1]{#1}
\newcommand{\newIII}[1]{#1}

\begin{document}

\title{Constraining the vertical structure of the Milky Way
rotation\\
by microlensing in a finite-width global disk model}

\authorrunning{J. Ja{\l}ocha, S. Sikora, {\L}. Bratek, M.
Kutschera }
\titlerunning{The disk thickness of the Milky Way.}
\author{
Joanna Ja{\l}ocha \inst{1}, Szymon Sikora \inst{2},
 {\L}ukasz Bratek
\inst{1}\fnmsep\thanks{\email{Lukasz.Bratek@ifj.edu.pl} },    \and
Marek Kutschera \inst{3}}
 \institute{
Institute of Nuclear Physics, Polish Academy of Sciences,
Radzikowskego 152, PL-31342 Krak\'{o}w, Poland \and Astronomical
Observatory, Jagellonian University, Orla 171, PL-30244
Krak{\'o}w, Poland
 \and
 Institute of
Physics, Jagellonian University,  Reymonta 4, PL-30059 Krak{\'o}w,
Poland }

\abstract{In this paper we model the vertical structure of mass
distribution of the Milky Way galaxy in the framework of a
finite-width global disk model. Assuming the Galactic rotation
curve only, we test inside the solar orbit the predictions of the
model for two measurable and unrelated to each other processes:
the gravitational microlensing that allows to fix the disk
width-scale by the best fit to measurements, and the vertical
gradient of rotation modelled in the quasi-circular orbits
approximation. The former is sensitive to the gravitating mass in
compact objects and the latter is sensitive to all kinds of
gravitating matter. The analysis points to a small width-scale of
the considered disks \new{and, at most, insignificant contribution
of non-baryonic dark mater in the solar circle. The predicted high
vertical gradient values in the rotation are consistent with the
gradient measurements.}}


\maketitle

\section{Introduction}

The Milky Way (\mw) is an example of a galaxy with high vertical
gradients of rotation measured at low altitudes above the
mid-plane \citep{2008ApJ...679.1288L}. It is interesting to see
the effect of the disk thickness on the gradient value prediction
and to compare it with the analogous prediction in an
infinitesimally thin global disk model
\citep{2010MNRAS.407.1689J}.

The gravitational microlensing phenomenon provides another
constraint on the mass distribution, independent of the vertical
gradient structure. In particular, the amount of mass seen through
the gravitational microlensing measurements inside the solar
orbit, was shown to be consistent with the dynamical mass
ascertained from the Galactic rotation and reduced by the gas
contribution undetectable by the microlensing
\citep{2012A&A...546A.126S}.

In the present paper we use microlensing measurements to constrain
the disk width and we compare the resulting vertical gradient
predictions with the gradient measurements. In our several
previous studies, e.g.,
\citet{2010MNRAS.407.1689J,2011MNRAS.412..331J}, we modelled
spiral galaxies in the approximation of an infinitesimally thin
disk. In this framework, we could easily obtain high values of the
vertical gradient of rotation, in accord with gradient
measurements. However, the model does not account for the vertical
structure in the mass distribution in the direct neighborhood of
the mid-plane (the $z=0$ vicinity). This structure can be modelled
by considering a finite-width disk
\new{with an assumed vertical profile of mass density}. A preliminary
step towards this we presented in \citet{2012A&A...546A.126S},
where a column mass density \new{of a disk with the exponential
vertical profile} was identified with the surface density of an
infinitesimally thin disk. \add{But this was a simple
de-projection which did not take into account the redistribution
of mass required to preserve the shape of rotation curve in the
Galactic center vicinity.}

In the next approximation considered in this paper, in
\secref{sec:DiskTheory} and \secref{sec:lensing1}, we find \new{by
iterations the volume mass densities of two example finite-width
disks}. These disks exactly account for the \new{\mw's tabulated}
rotation curve \new{and have their width-scales constrained by the
gravitational microlensing measurements. In \secref{sec:gradient}
we compare the predictions of the two models with the measurements
of the vertical gradient of rotation}.

\section{\label{sec:DiskTheory}A finite-width disk model}

We assume the following factorized form of a volume mass density
$\rho(r,z)=\sigma(r)f(z)$, with factors normalized so that
$\sigma(r)$ is the column mass density and the vertical profile
$f(z)$ is integrable to \new{unity:
$\int_{-\infty}^{\infty}f(z)\ud{z}=1$}.  In this paper we consider
the exponential vertical profile (for its simplicity) and the
\mexhat profile \new{(for its being frequently used in star count
models motivated by \citet{1942ApJ....95..329S} analysis)}:
\begin{equation}\label{eq:profiles}
f(z)=\frac{1}{2h'}\,\mathrm{exp}\br{-\frac{\abs{z}}{h'}},\qquad
f(z)=\frac{1}{2h}\,\mathrm{sech}^2\br{\frac{z}{h}}.\end{equation}
The width-scale parameters $h'$ and $h$ can be related one to
another by equating their effective disk thicknesses: $2\,h'$ and
$\approx1.49\,h$, respectively, \new{defined by the ``$1/\exp{}$
criterion''}.\footnote{\new{The ``$1/\exp{}$ criterion'' we
formulate as follows: the effective width-scale $h'$ of a vertical
profile $f(z)$ integrable to unity is defined by comparing its
mass inside a shell $|z|<h'$ with that of an exponential one with
a scale-width $h'$: $\int\limits_{-h'}^{h'}f(z)\ud{z}\equiv
\int\limits_{-h'}^{h'}\frac{\exp{-|z|/h'}}{2h'}\ud{z}=1-\frac{1}{e}$.
In particular, for the \mexhat profile we obtain
$\tanh\br{h/h'}=1-\frac{1}{e}$ or $2\,h'\approx1.49\,h$.}}

The gravitational potential at a point $\vec{X}=[R,0,Z]$
associated with a mass element
$\ud{m}=\rho(r,z)\,r\ud{r}\,\ud{\phi}\,\ud{z}$ located at another
point $\vec{Y}=[r\cos{\phi},r\sin{\phi},z]$, is $
\frac{-G\ud{m}}{\abs{\vec{Y}-\vec{X}}}$, where
$\abs{\vec{Y}-\vec{X}}=\sqrt{R^2+(z-Z)^2+r^2-2rR\cos{\phi}}$. On
substituting $\phi=2\alpha-\pi$, using the symmetry
$\phi\to-\phi$, integrating over $\alpha\in\br{0,\pi/2}$ and using
the definition of of the elliptic integral of the first kind $K$
\citep{Ryzhik}, one arrives at an expression for the total
potential $\Psi(R,Z)$ at $\vec{X}$: $\Psi(R,Z)=-\,4\,G
\int\limits_{0}^{\infty}\!\ud{r}\!\int\limits_{-\infty}^{\infty}\!\ud{z}\,
\frac{r\,K\br{\sqrt{\frac{4\,r\,R}{(R+r)^2+(z-Z)^2}}}\,\rho(r,z)}{\sqrt{(R+r)^2+(z-Z)^2}}$.
\rem{where $k=\sqrt{\frac{4\,r\,R}{(R+r)^2+(z-Z)^2}}$.} With the
help of the identity
$K'(k)=\frac{E(k)}{k\,(1-k^2)}-\frac{K(k)}{k}$ \citep{Ryzhik}, it
can be shown that $ R\,\partial_R\Psi(R,Z)=2\,G
\int\limits_{0}^{\infty}\ud{r}\int\limits_{-\infty}^{\infty}\ud{z}
\,\rho(r,z)\,J(r,R,z-Z)$. Function $J$ is an integration kernel
\add{that will frequently appear later}:
$$J(r,R,\zeta)=\frac{r\cdot\br{K(\kappa)-\frac{r^2-R^2+\zeta^2}{(r-R)^2+\zeta^2}\,E(\kappa)}}{\sqrt{(R+r)^2+\zeta^2}}
,\quad \kappa\!=\!\sqrt{\frac{4rR}{(R+r)^2+\zeta^2}}.$$ The
expression for $R\,\partial_R\Psi(R,Z)$ can be used to compute the
rotation velocity $v_{\phi}(R,Z)$ in the quasi-circular orbits
approximation studied in \citet{2010MNRAS.407.1689J}. In this
approximation, $v^2_{\phi}(R,Z)=R\,\partial_R\Psi$, and the
resulting vertical gradient of rotation is
$\partial_Zv_{\phi}(R,Z)$. The latter quantity would involve
differentiation of the kernel $J$ under the integration sign.
However, since $\rho(r,z)$ is of the form $\sigma(r)f(z)$, with
$f(z)$ being known in an analytic form, falling off fast enough as
$\abs{z}\to\infty$, and satisfying the reflection symmetry
$f(z)=f(-z)$, it is more convenient to integrate by
parts.\rem{\footnote{\rem{It is convenient to assume that
$\int_{-\infty}^{\infty}f(z)\ud{z}=1$, then $\sigma(r)$ has the
interpretation of a column mass density. For example, if
$\rho(r,z)=\sigma(r)\frac{\exp{-\abs{z}/h}}{2h}$, then
$\partial_{z}\rho(r,z)=-\frac{1}{h}\rho(r,z)$ for $z>0$.}}} Noting
that $\partial_zJ=-\partial_ZJ$, \rem{and using the symmetry of
$f$,} we are led to the following expressions for the circular
velocity and its vertical gradient (by the assumed reflection
symmetry of $f$, the integration has been restricted to $z>0$):
\begin{eqnarray} v^2_{\phi}(R,Z)&=&2G\!
\int\limits_{0}^{\infty}\!\!\ud{r}\!\!\int\limits_{0}^{\infty}\!\!\ud{z}\,
\rho(r,z)\cdot\nonumber\\&&\phantom{XXXXXXX}\cdot\sq{J(r,R,z-Z)+J(r,R,z+Z)},\label{eqn:rotation_thindisk}
\\ \partial_Zv_{\phi}(R,Z)&=&\frac{G}{v_{\phi}(R,Z)}
\int\limits_{0}^{\infty}\!\!\ud{r}\!\!\int\limits_{0}^{\infty}\!\ud{z}
\,\partial_z\rho(r,z)\cdot\nonumber\\&&\phantom{XXXXXXX}\cdot\sq{J(r,R,z-Z)-J(r,R,z+Z)}.\label{eqn:gradient_thindisk}
\end{eqnarray} \new{(The derivatives of $J$ have been elliminated from \eqref{eqn:gradient_thindisk}
by means of an integration by parts.)} It should be stressed that
these two expressions are valid in the quasi-circular orbits
approximation only.

The above integral expressions are particularly suited for the
exponential vertical profile, in which case they reduce to
$v^2_{\phi}(R,Z)=I_{-}+I_{+}$ and
$\partial_Zv_{\phi}(R,Z)=\frac{I_{+}-I_{-}}{2h\sqrt{I_{-}+I_{+}}}$,
where $I_{-}$ and $I_{+}$ are appropriate integrals involving
$J(r,R,z-Z)$ and $J(r,R,z+Z)$, respectively.

\new{For a more explicit derivation of the above results, the reader
is referred to the Appendix.}

\subsection{\label{sec:rhodetermination}Determining $\rho(r,z)$
from the rotation curve by iterations}

\begin{figure}
\centering
\includegraphics[trim = 2.65mm 12.2mm 0mm 0mm, clip,width=0.5\textwidth]{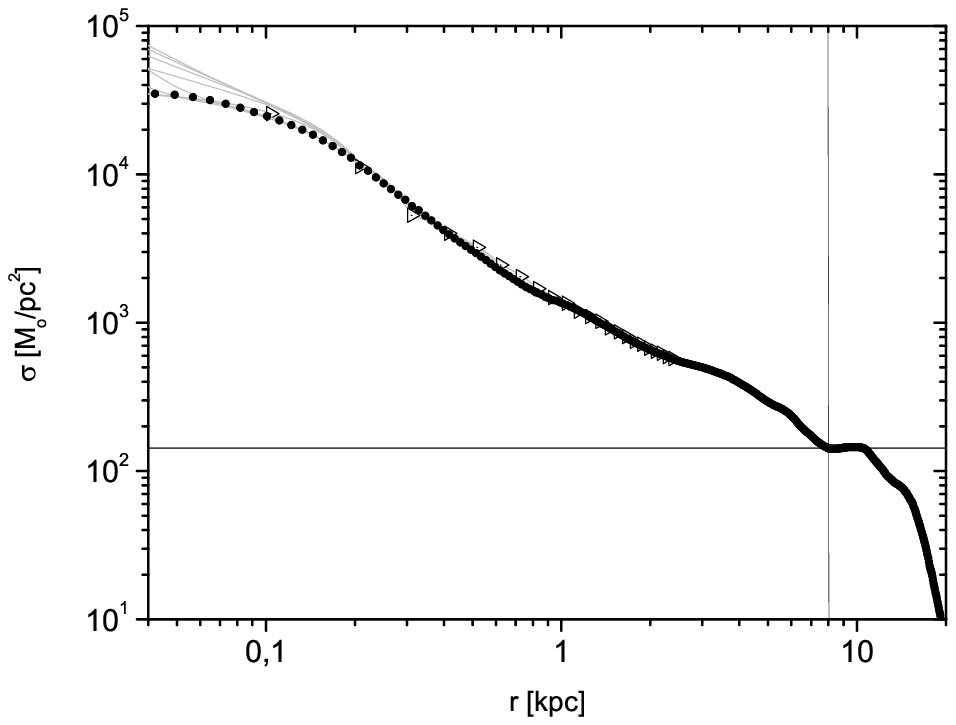}
\includegraphics[trim = 0mm 5mm 0mm 6.9mm, clip,width=0.5\textwidth]{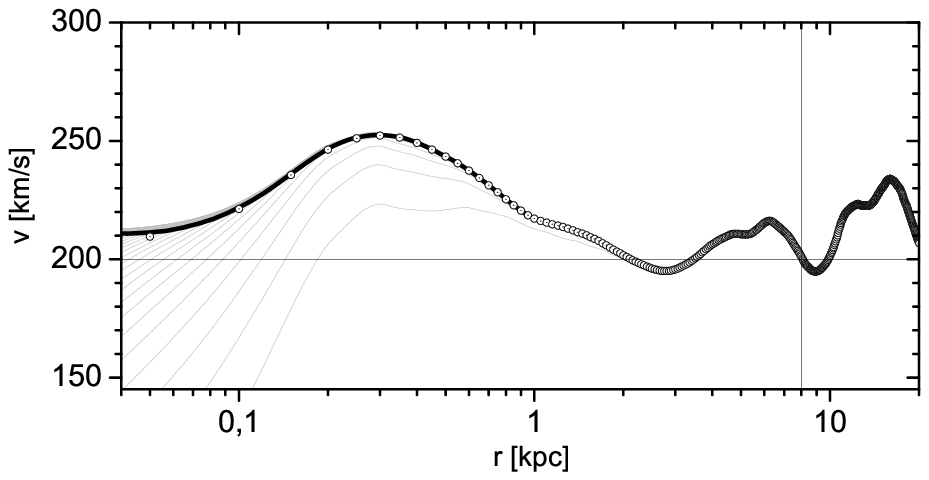}
\caption{\label{fig:dodatkowy1}
\new{
Column mass densities and the corresponding rotation curves of
finite-width disks with a \mexhat vertical profile ($h=117\pc$),
shown at several steps of the iteration method of
\secref{sec:rhodetermination}, starting from the surface density
of an infinitesimally thin disk.\newline \textbf{Top panel:}
\crv{thin gray lines} -- column mass densities at various
iteration steps, \crv{triangles} -- the surface mass density of
the infinitesimally thin disk model (the starting point of the
iteration), \crv{solid circles} -- the \mw column mass density in
the final iteration step (it gives the interpolating rotation
curve in the bottom panel). \newline \textbf{Bottom panel:}
\crv{open circles} -- \mw rotation curve points from
\citet{1999ApJ...523..136S}, \crv{thin gray lines} -- rotation
curves at various iteration steps, and \crv{thick line} --
rotation curve in the final iteration step (it strictly overlaps
with a spline interpolation of the \mw rotation curve points).}}
\end{figure}

We use \new{a smoothed-out \mw} rotation curve from
\citet{1999ApJ...523..136S}, \new{which adopts Galactic constants
$\rsun=8\,\kpc$ and $\vsun=200\,\kms$.} Its inner part, inside
solar circle, which is of interest in our paper, \new{is
relatively well determined and was obtained by simple averaging of
various CO and HI} tangent velocity data. \rem{combined in
\citep{1985ApJ...295..422C}} The uncertainty lies mostly in the
velocity parameter of the standard of rest at the Sun position,
\new{$\vsun$}, and the radius of solar orbit
\new{$\rsun$}. For larger radii, outside the region of our
interest, the rotation curve is less certain and even model
dependent. \new{Furthermore, as follows from the analysis in
\citet{2008MNRAS.391.1373B}, the uncertainty in the external
rotation curve implies some uncertainty in the internal mass
determination due to a backward-interaction characteristic of
flattened mass distributions. However, in
\citet{2012A&A...546A.126S} we found this influence to be marginal
for the purpose of the present study. More important is to reduce
the uncontrolled numerical errors arising due to the presence of
singular kernels in integrals \eqref{eqn:rotation_thindisk} and
\eqref{eqn:gradient_thindisk}. To achieve this, we apply a cubic
spline interpolation to the rotation points. This way we obtain a
continuously differentiable interpolating rotation curve (see
\figref{fig:dodatkowy1}).}

A primary approximation to the column mass density, $\sigma_0(R)$,
we obtain from the \new{interpolating} rotation curve in the
infinitesimally thin global disk model, by iterations akin to
those described in \citet{2008ApJ...679..373J}, with the help of
the following integral transforms which are mutual inverses of
each other \citep{2012A&A...546A.126S}
\begin{equation}\label{dysk_sigma}
\sigma_0\br{R}=\frac{1}{2\pi^2G}\int\limits_{0}^{\infty} \sq{
\frac{K\!\sq{k(\chi)}}{1+\chi}-
\frac{E\!\sq{k(\chi)}}{1-\chi}}\frac{v^2\!\br{R\,\chi}}{R\,\chi}\,\ud{\chi}.\end{equation}
and
\begin{equation}\label{dysk_rotacja}\frac{v^2\br{R}}{R}=2G\int\limits_{0}^{\infty}
\sq{ \frac{K\!\sq{k(\chi)}}{1+\chi}+
\frac{E\!\sq{k(\chi)}}{1-\chi}}\sigma_0\!\br{R\,\chi}\chi\,\ud{\chi},
\end{equation} with $k(\chi)=\frac{2\sqrt{\chi}}{1+\chi}$ in both
cases. The transforms relate the rotation law in the Galactic
mid-plane to the surface mass density in that plane. Here, $R$ is
the radial variable in the disk plane, $K$ and $E$ are complete
elliptic integrals of the first and second kind as defined in
\citep{Ryzhik}, and $\chi$ is a dimensionless integration
variable. The surface mass density $\sigma_0(r)$ obtained this way
gives a primary approximation $\rho_0(r,z)=\sigma_0(r)f(z)$ to the
volume mass density, which should be close to the mass density in
a finite-width disk (the results of \citet{2012A&A...546A.126S}
were based on this approximation). But evaluating integral
\eqref{eqn:rotation_thindisk} with $\rho_0(r,z)$ substituted for
$\rho(r,z)$, gives a $v^2_{\phi,0}(R,0)$ different from $v_c^2(R)$
by a small amount $\Delta v^2_{\phi,0}(R)$ measuring the
discrepancy between the predicted and the observed rotation. Now,
by inserting the $\Delta v_{\phi,0}^2(R)$ in place of $v^2$ in
\eqref{dysk_sigma}, we obtain a correction $\Delta\sigma_0(R)$ to
the column mass density. Hence, the surface mass density in the
next approximation is\footnote{The free parameter $\lambda$, can
be used to control the rate of convergence of the iteration
process. We used $\lambda=\frac{2}{3}$ instead of $1$.}
$\sigma_1(R)=\sigma_0(R)+\lambda\cdot\Delta\sigma_0(R)$ and gives
rise to a corrected volume mass density
$\rho_1(r,z)=\sigma_1(r)f(z)$, \new{which in turn, from
\eqref{eqn:rotation_thindisk}, gives the corresponding corrected
rotation curve $v_{\phi,1}(R,0)$. Next, we shift all indices by
$+1$ and repeat the previous step.} This correction process can be
continued recursively until it converges to the desired density
profile $\rho(r,z)$ for which the discrepancy $\Delta
v^2_{\phi,j}$ becomes negligible. \new{The resulting recursion
sequence of volume densities
$\rho_j(R,Z)=\rho_{j-1}(R,Z)+\lambda\cdot
f(Z)\Delta\sigma_{j-1}(R)$ is quickly converging to a limit
$\rho(R,Z)=\lim\limits_{j\to\infty}\rho_j(R,Z)$ which, on
substituting to \eqref{eqn:rotation_thindisk}, yields a curve
$v_{\phi}(R,0)$ that nicely overlaps with the interpolating
rotation curve. The iteration process is illustrated in
\figref{fig:dodatkowy1}.}

In \secref{sec:lensing1} we use this method of finding column mass
densities $\sigma(R)$ corresponding to various vertical profiles
$f(z)$. They all give rise to rotation laws that overlap in the
mid-plane  with the \mw rotation curve.

\section{\label{sec:lensing1}The gravitational microlensing}

The phenomenon of the gravitational light deflection in point mass
fields, called the gravitational microlensing, may be used to
tracing mass distribution. This can be achieved by inferring the
amount of mass in compact objects scattered along various lines of
sight joining the observer and remote sources of light
\add{distributed} in the vicinity of the Galactic center. For it
is not directly linked to the Galaxy dynamics, the gravitational
microlensing provides an independent tool useful in testing
Galactic models.

The original idea comes from \citet{1986ApJ...304....1P}, who
suggested that the gravitational lensing can be used to answer the
question of whether or not the spheroidal component of the Galaxy
could be dominated by massive compact objects. Observations
towards Great and Small Magellanic Clouds, e.g.,
\citet{2000ApJ...542..281A,2011MNRAS.413..493W}, indicate that the
contribution from such objects cannot be dominating in the
standard three-component galactic model. On the other hand, it
turns out that the microlensing method \add{can be used also} to
estimate the distribution of compact objects in the direction
towards the Galactic center. Such measurements can help to improve
models of the galactic interior, see e.g.,
\citet{2004ApJ...601L.155B}. It should be pointed out that in most
of these models, the dark matter component becomes necessary
\add{only} beyond some particular distance from the Galactic
center. For example, in the model by \citet{2002MNRAS.330..591B},
the dark matter halo is irrelevant for distances smaller than
$5\,\mathrm{kpc}$ from the center.

The aim of the present work is to examine in the context of the
microlensing observations, a galactic model in which the dynamics
is dominated by baryonic matter distributed in direct neighborhood
of the Galactic mid-plane.

\subsection{The optical depth}

The most important quantity to be determined in the microlensing
method is the optical depth $\tau$. It is defined as the
probability of finding a compact object (a lens) on the line of
sight between the observer and the source of light, when a lens is
located within its Einstein radius
$$R_E=\sqrt{\frac{4GM}{c^2}\,\frac{D_L\,(D_S-D_L)}{D_S}} $$ on a
plane perpendicular to the line of sight. Here, $M$ denotes the
lens mass, $D_S$ is the distance from the observer to the source,
and $D_L$ is the distance between the observer and the lens. In
this particular configuration, a double image of the source is
produced each time a lens passes between the source and the
observer. Although the two images cannot be usually resolved
\add{due} to very small deflection angles, their appearance
\add{can still be} detected indirectly by measuring the associated
\add{image magnification}. A microlensing event of this kind is
agreed to have been occurred when \add{the magnification} exceeds
a threshold value of $\mu=1.34$. Because the probability of
microlensing events is very low (of the order of $10^{-6}$), a
great number of sources (of about a million) must be monitored
during a period of few years \add{so as to} the optical depth
could be determined. A detailed discussion of such observations
and their theoretical description can be found in
\citet{2010GReGr..42.2047M}.

The principle result of the microlensing theory is the following
integral that relates the mass density of compact objects
$\rho(\vec{x})$ and the optical depth $\tau$:
\begin{equation}\label{eq:tau} \tau=\frac{4\pi
G}{c^2}\int_{0}^{D_S}\frac{D_L\,(D_S-D_L)}{D_S}\,\rho(\vec{x}(D_L))\,\mathrm{d}D_L
\,. \end{equation} Although it is well known, we re-derived this
formula for completeness in our previous paper
\citep{2012A&A...546A.126S}. The integration in \eqref{eq:tau} is
carried out along a given line of sight, between the observer's
position $\vec{x}_{\odot}=[\rsun,0,0]$ and the source of light
located at $\vec{x}_\otimes=\vec{x}_{\odot}+(1+\chi)\rsun\,[-\cos
b\cos l,-\cos b\sin l,\sin b]$. The angle $l$ is the galactic
longitude, $b$ is the galactic latitude and $\chi$ is a
dimensionless distance parameter, such that $D_S=\rsun\,(1+\chi)$.
Using the following parameterization of the line of sight
$\vec{x}(s)=\vec{x}_{\odot}+s\,(\vec{x}_\otimes-\vec{x}_{\odot})$,
equation \eqref{eq:tau} can be rewritten \add{such that} $\tau$
becomes an explicit function of the source's position $(l,b,\chi)$
\citep{2012A&A...546A.126S}:
\begin{equation}\label{optical_depth_blchi}
\tau(l,b,\chi)=\frac{4\pi
G\rsun^2}{c^2}\int_{0}^{1}(1+\chi)^2\,s(1-s)\,
\rho(r(s),z(s))\,\mathrm{d}s\,, \end{equation} where
$r(s)=\rsun\,\sqrt{1+s\,(1+\chi)\cos b\,\left[s\,(1+\chi)\cos
b-2\cos l \right]}$ and $z(s)=\rsun\,s\,(1+\chi)\sin b$. Given a
density distribution $\rho(r,z)$, the above formula enables one to
calculate the corresponding optical depth.

\subsection{The data}\label{sec:data_points}

The model optical depth, calculated with the help of expression
\eqref{optical_depth_blchi}, \add{must} be compared with the
observations. For that purpose, we use data collected by several
leading collaborations, in particular: MACHO
\citep{2005ApJ...631..879P}, EROS \citep{2006A&A...454..185H},
OGLE \citep{2006ApJ...636..240S}, and MOA
\citep{2003ApJ...591..204S}. These data were collected and
analyzed in detail in a review \citep{2010GReGr..42.2047M}.

\add{Later, we} restrict our analysis to the bright stars
subsample. This is a commonly used strategy \add{in minimizing}
blending \add{processes that affect} the optical depth results. A
discussion of the blending effect can be found in
\citet{1997A&A...321..424A} and \citet{2007MNRAS.380..805S}. The
data we use is represented as a function of latitude $\tau(b)$,
and for each $b$ the optical depth \add{has been} averaged over
the longitudinal angle in the interval $l\in(-5^\circ,5^\circ)$.
The resulting function $\tau(b)$ allows us to study the vertical
structure of the Galaxy.

\subsection{Previous results in view of the present study}

In our previous paper \citep{2012A&A...546A.126S}, we used a
surface mass density $\sigma(r)$, which accounts for the Galactic
rotation curve in the infinitesimally thin disk model, to obtain a
volume mass density $\rho(r,z)$ corresponding to $\sigma(r)$,
assuming the standard exponential vertical profile
$\rho(r,z)=\rho(r,0)\,\exp{-|z|/h'}$. Following the EROS
collaboration, e.g., \citet{2001A&A...373..126D}, we set the value
of the width-scale parameter to be $h'=325\pc$. With such obtained
$\rho(r,z)$, we showed that the resulting optical depth was
consistent with the observational data at a reasonable confidence
level. In addition, we investigated several issues which could
possibly influence the optical depth, among them: the uncertainty
of the solar Galacto-centric distance $\rsun$, the problem of a
precise determination of the rotation curve, the difference
between a single exponential vertical profile and a double
exponential profile, and the structure of the central bulge. We
pointed out that the optical depth uncertainty connected with each
of these factors was relatively small, and could not spoil the
consistency between the model predictions and the observations.

The purpose of the present work is to check the microlensing
optical depth predictions within a finite-width disk model
framework, assuming a spatial mass distribution derived directly
from the \new{interpolating} rotation curve, as it was described
in \secref{sec:rhodetermination}. In this model the disk thickness
is crucial and it affects the distribution of column mass density
\add{which can change while keeping the predicted rotation
unchanged and identical with the interpolating rotation curve of}
\secref{sec:rhodetermination}.  As it was mentioned earlier, we
assume a volume mass density of the form
$\rho(r,z)=\sigma(r)f(z)$, where $\sigma(r)$ is the column mass
density and $f(z)$ is either the \add{normalized \mexhat or
exponential vertical profiles given in} \eqref{eq:profiles}. Each
of these profiles is defined by its own characteristic
width-scale, $h$ or $h'$, which are free parameters. Their optimal
values can be constrained with the help of microlensing
measurements, as it is done in the next subsection.

\subsection{\label{sec:microlensing}
Microlensing in a finite-width disk model}

The sources of light observed in the microlensing events are
randomly distributed in the vicinity of the Galactic center. This
requires some averaging in the latitudinal angle $b$. Hence, the
optical depth observable (as a function of $b$) should be
understood as a moving average. It is frequently assumed that the
sources are located on the symmetry axis. We adopt this
simplification in calculating $\tau(b)$ by substituting $l=0$ and
$\chi=0$ in \eqref{optical_depth_blchi}.

\begin{figure}
\centering
\includegraphics[angle=-90,width=0.5\textwidth]{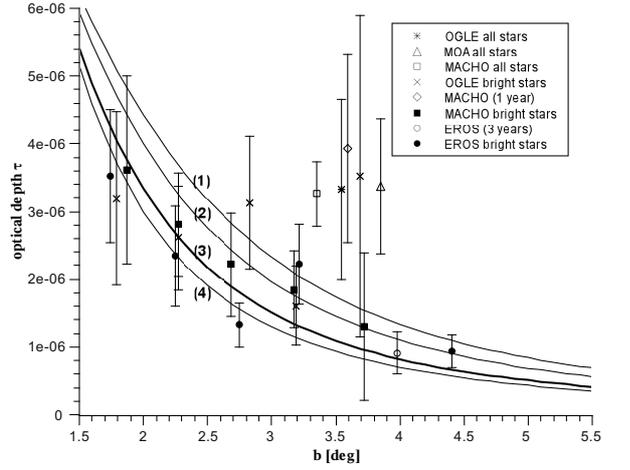}
\caption{\label{fig:micro1} \crv{solid lines}: the optical depth
$\tau(b)$ in the finite-width disk model for the \mexhat
\add{vertical} profile, with the width-scales: (1) $h=180\pc$, (2)
$h=150\pc$, (3) $h=117\pc$ /\emph{the best fit}/, and (4)
$h=100\pc$. The points with error bars represent the measurement
data collected by several collaborations (references in the
text).}
\end{figure}

\begin{figure}
\centering
\includegraphics[angle=-90,width=0.5\textwidth]{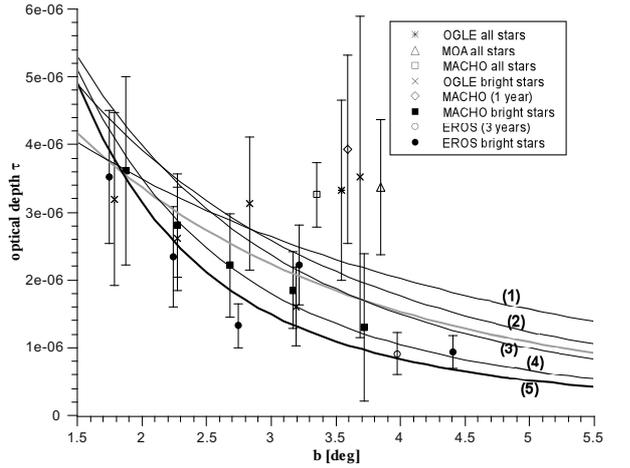}
\caption{\label{fig:micro2} \crv{solid lines}: the optical depth
$\tau(b)$ in the finite-width disk model for the exponential
\add{vertical} profile with the width-scales:  (1) $h'=325\pc$,
(2) $h'=200\pc$, (3) $h'=150\pc$, (4) $h'=100\pc$, (5) $h'=88\pc$
/\emph{the best fit}/. For comparison, the \crv{gray thick line}
represents the $\tau(b)$ in the infinitesimally thin disk model
for which the equivalent volume mass density was obtained assuming
the exponential vertical profile with $h'=325\pc$. The points with
error bars represent measurements data collected by several
collaborations (references in the text).}
\end{figure}

Our idea of determining $\tau(b)$ is simple. For a fixed
width-scale \new{$h$ or $h'$}, we obtain a volume mass density
$\rho(r,z)$ from the Galactic rotation curve, as described in
detail in \secref{sec:rhodetermination}. With this $\rho(r,z)$ we
calculate the optical depth $\tau(b)$ from
\eqref{optical_depth_blchi}, substituting $l=0$ and $\chi=0$. The
results for the vertical density profiles \eqref{eq:profiles} are
shown in \figref{fig:micro1}
and \figref{fig:micro2},
respectively. The curves, corresponding to different
\add{width-scale values}, are shown together with the observed
data points described in \secref{sec:data_points}. We remind that
we are interested in the bright stars sub-sample only.

To measure the accuracy of the fitting curves \add{obtained} for
various width-scales, we calculate the reduced \Q{chi squared},
$\chi_{red}^2$, that is, the \Q{chi squared} divided by the number
of the degrees of freedom (the fit residuals from the bright stars
sub-sample were taken into account).
\begin{figure}
   \centering
      \includegraphics[angle=-90,width=0.5\textwidth]{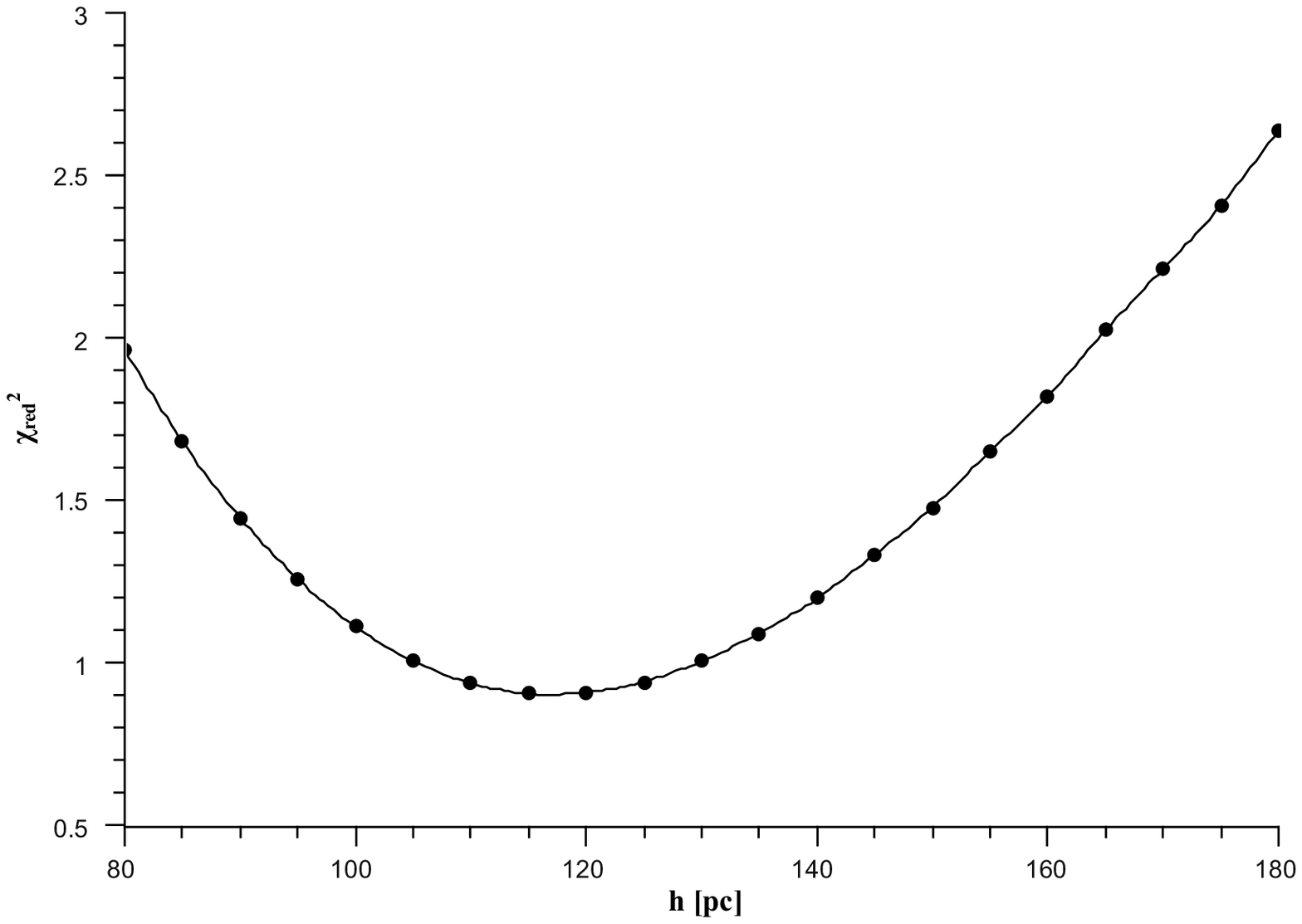}
      \includegraphics[angle=-90,width=0.5\textwidth]{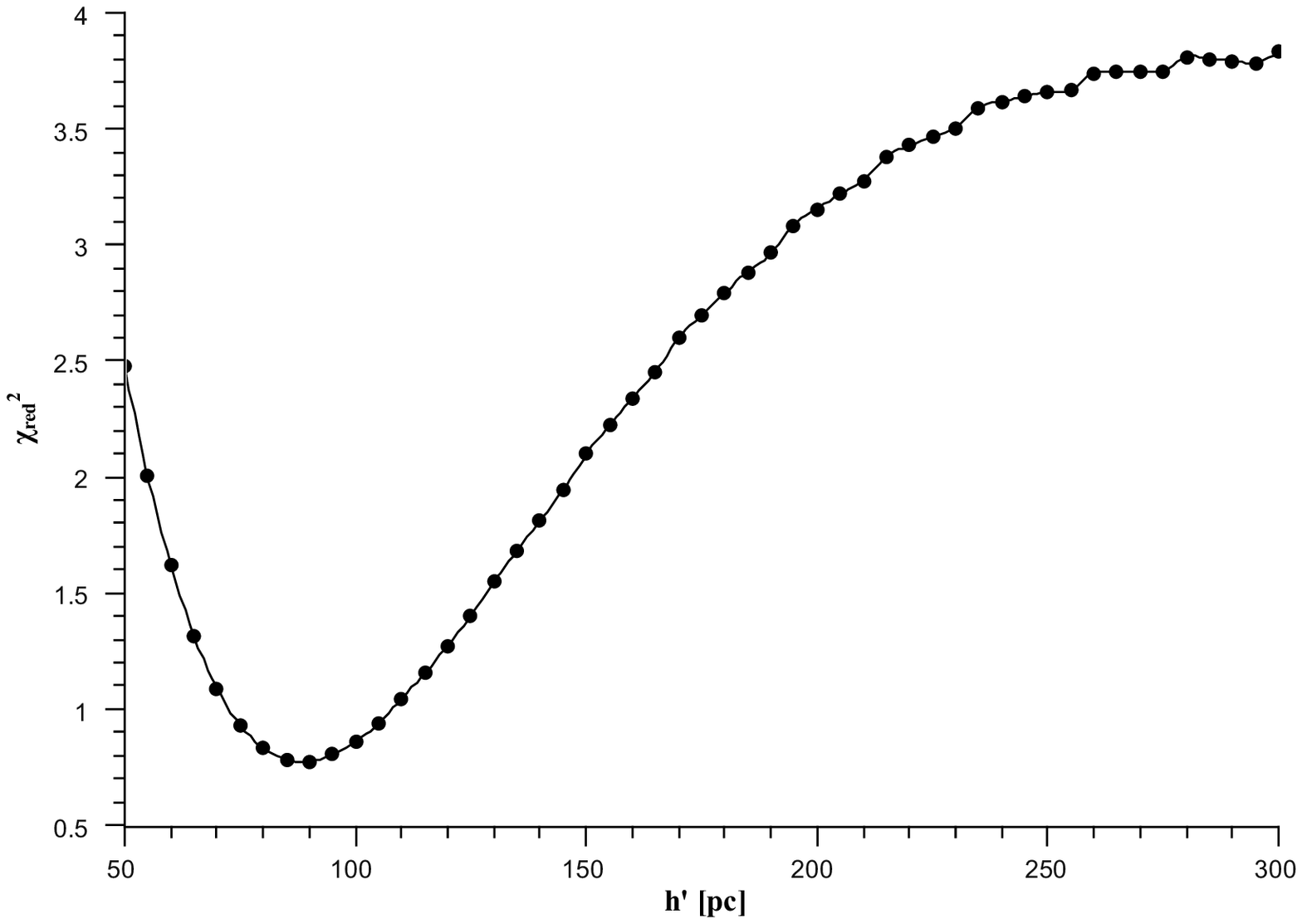}
      \caption{\label{fig:micro3}
The reduced \Q{chi squared}, $\chi_{red}^2$, as a function of the
width-scale parameter: $h$ (for the \mexhat \add{vertical} profile
\crv{top panel}) or $h'$ (for the exponential \add{vertical}
profile \crv{bottom panel}). The values of the width-scales that
minimize the $\chi_{red}^2$ are: $h=117\pc$ and $h'=88\pc$,
respectively.}
  \end{figure}
The $\chi_{red}^2$ are plotted in \figref{fig:micro3} for several
width-scales $h$ (black dots in the top panel) and $h'$ (black
dots in the bottom panel). \add{By cubic-spline-interpolating the
$\chi_{red}^2$, we may regard it} as a smooth function of the
width-scale. This allows us to determine the optimum width-scale
at the minimum of each $\chi_{red}^2$. The width-scales
\add{obtained} this way are \add{$h=117\pc$ and $h'=88\pc$,
respectively, for the \mexhat and the exponential vertical
profiles}.
\begin{figure}
   \centering
      \includegraphics[angle=-90,width=0.5\textwidth]{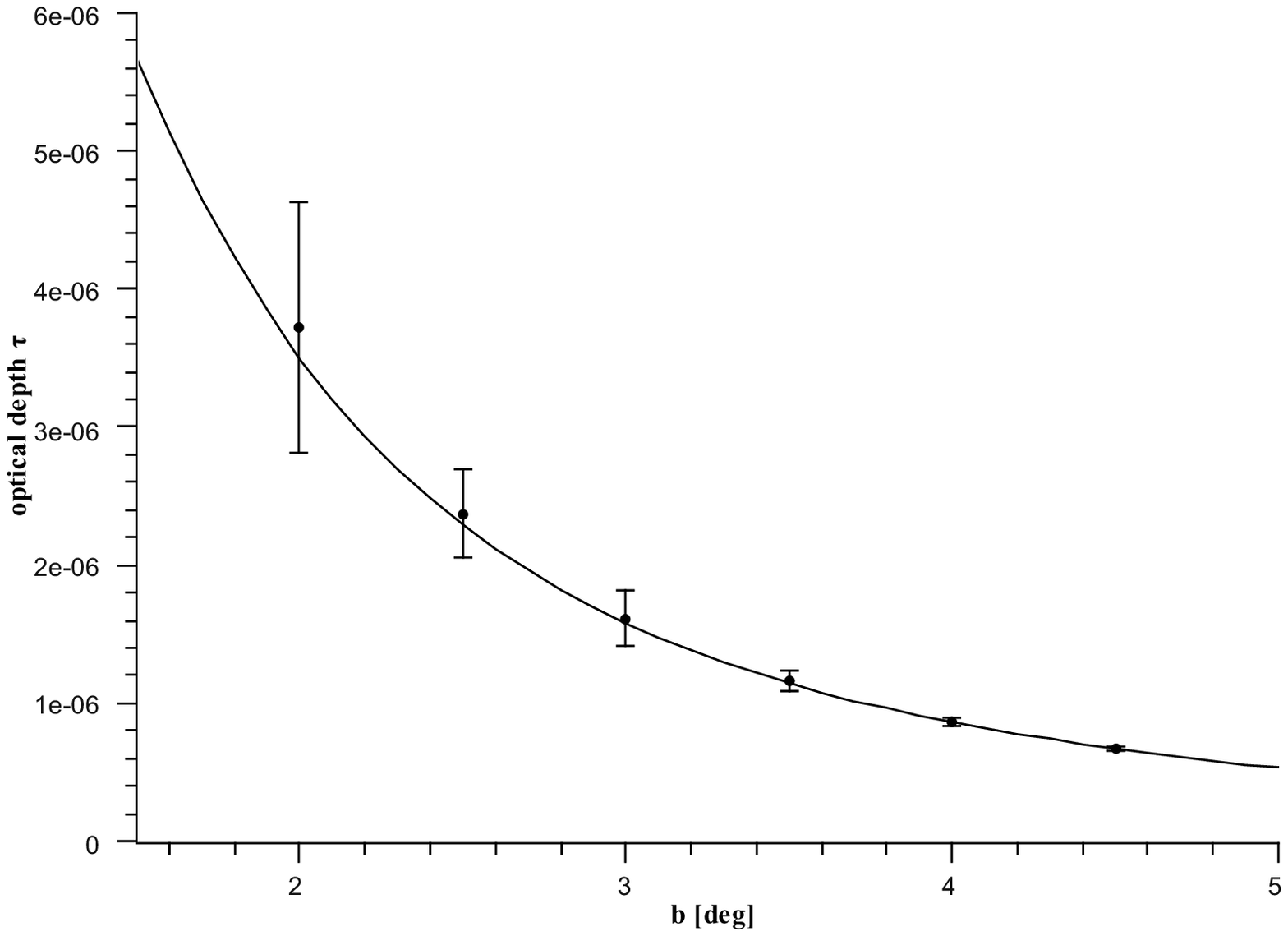}
      \includegraphics[angle=-90,width=0.5\textwidth]{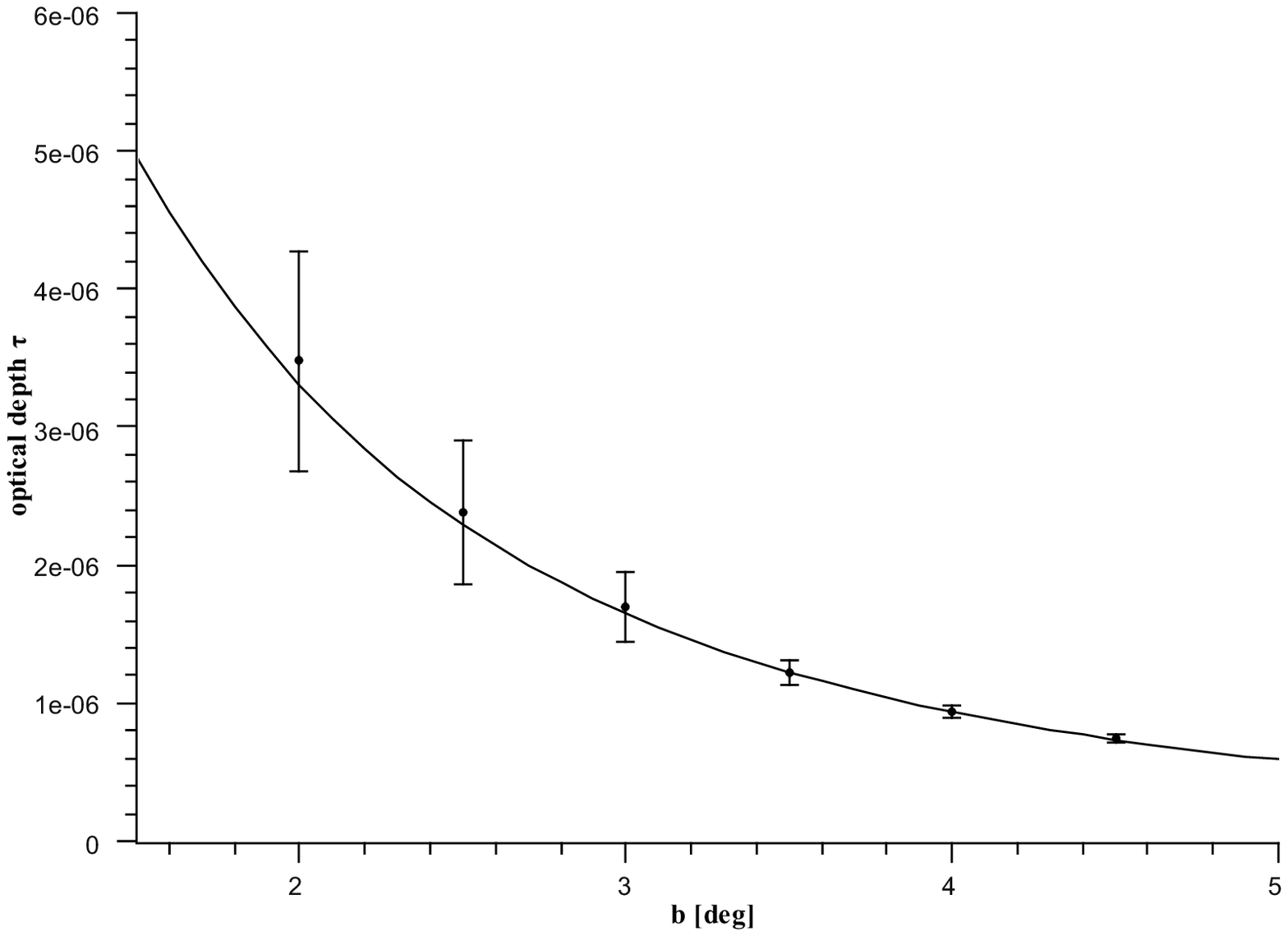}
      \caption{\label{fig:micro4}
The results of a Monte Carlo simulation: the mean optical depth
and its standard deviation for a sample of sources chosen randomly
with a probability distribution proportional to $\rho(r,z)$.
\crv{points+error bars}. To compare with, the \crv{solid line}
shows the optical depth calculated with the help of the integral
\eqref{optical_depth_blchi} with $l=0$ and $\chi=0$. The \crv{top
panel} shows the results for the \mexhat \add{vertical} profile
with $h=117\pc$, and the \crv{bottom panel} those for the
exponential \add{vertical} profile with $h'=88\pc$.}
  \end{figure}

Finally, we need to verify whether the approximation of sources
aligned to the symmetry axis is justified. To this end we
performed a Monte Carlo simulation. We chose randomly a number of
$n=10^4$ pairs $(l,\chi)$ in the range $\chi\in(-0.125,0.125)$
(corresponding to $\pm 1\,\kpc$) and $l\in(-5^\circ,5^\circ)$ for
each fixed latitude $b$, with the probability weight directly
proportional to $\rho(r,z)$, simultaneously finding the
corresponding $\tau(l,b,\chi)$ as defined in
\eqref{optical_depth_blchi}. The resulting mean optical depth  and
its standard deviation are shown as functions of $b$ in
\figref{fig:micro4}. The mean value is close to $\tau(0,b,0)$
(solid line), which proves the approximation \add{to be} quite
accurate.

\section{\label{sec:gradient}The vertical gradient of rotation}

In \citet{2010MNRAS.407.1689J} we modelled the vertical gradient
of \mw's rotation in the framework of infinitesimally thin disk
model. We compared the predicted high absolute gradient values
with the gradient measurements by \citet{2008ApJ...679.1288L}, and
found them to be consistent with each other. In what follows, to
see the influence of the {vertical structure of mass distribution
on the gradient value and its behavior}, we will repeat these
studies in a more accurate model accounting for a finite disk
thickness.

The microlensing results of \secref{sec:microlensing} imply a
width-scale of $117\pc$ for the disk with the \mexhat vertical
profile, and $88\pc$ for that with the exponential vertical
profile. With the corresponding mass distributions
\add{substituted in \eqref{eqn:gradient_thindisk}, we find our
prediction for the vertical gradient of rotation in the
rectangular measurement region of \citet{2008ApJ...679.1288L}:
$r\in\br{3,8}\,\kpc$ and $z\in\br{0,1.8}$.} Our predictions for
the gradient in this region are shown in \figref{fig:grad}, where
they were also compared with the predictions of the
infinitesimally thin disk model \new{and with those of
finite-width disks with the exponential vertical profile}.

\begin{figure}   \centering \includegraphics[trim = 0mm
6mm 0mm 0mm, clip,width=0.5\textwidth]{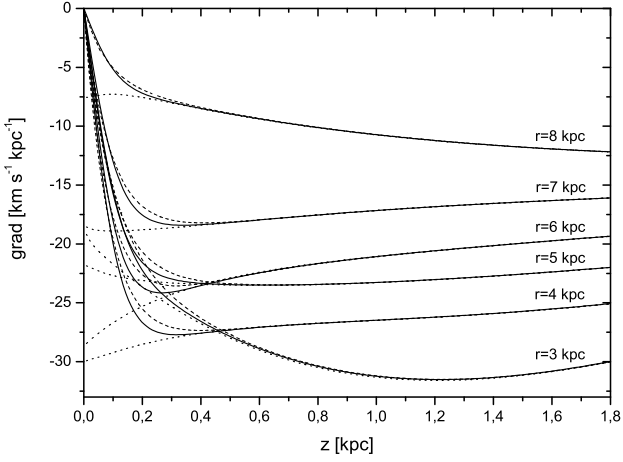}
\caption{\label{fig:grad}The vertical gradient of rotation \add{in
finite-width disk models,} as a function of altitude above the
mid-plane $z=0$, shown at various radii: \crv{solid lines} -- the
gradient for a disk with the \mexhat vertical profile ($h =
117\pc$), and \crv{dashed lines} -- the gradient for a disk with
the exponential vertical profile ($h' = 88\pc$). \add{To emphasize
the universal gradient behavior at higher altitudes for various
disks of the same mass, the results for an infinitesimally thin
disk model (with the surface mass density identified with the
column mass density of the disk with the \mexhat vertical profile)
was shown -- \crv{dotted lines}.}}  \end{figure}

\add{For all finite-width disks, the gradient falls off} to $0$ at
$z=0$, owing to the smoothness and $z$-reflection symmetry of the
mass distribution (then, the gradient is $z$-antisymmetric), while
for the infinitesimally thin disk with a mass distribution
singular on the symmetry plane, the gradient attains a finite and
large value at $z=0$ \new{(this requires the gradient line to be
discontinuous at $z=0$). For low altitudes, the gradient behavior
is dependent on the particular structure of mass distribution
decided by the parameter $h$ which introduces a characteristic
length that scales: the altitudinal extent of the \Q{turn-overs}
(seen in the gradient lines with $h>0$), their local minimum
positions, and the degree of their curvatures.} \add{Another
feature evident from \figref{fig:grad}, is the fact that there is
little difference between disk models with the exponential and
with the \mexhat vertical profile.}

\begin{figure}
\centering
\includegraphics[trim
= 0mm 6mm 0mm 0mm, clip,width=0.5\textwidth]{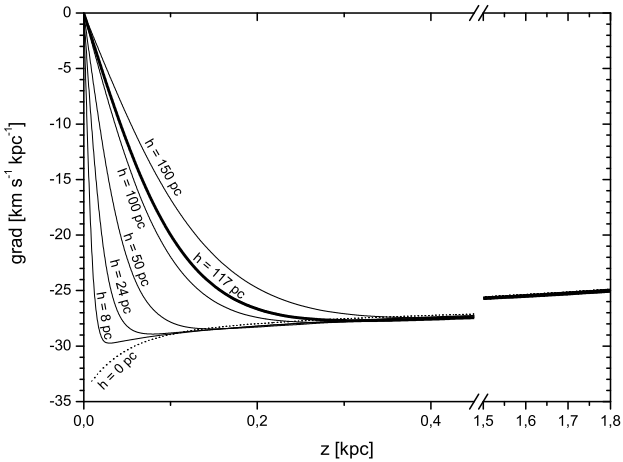}
\caption{\label{fig:grad3} \new{The vertical gradient of rotation
in a finite-width disk model with the \mexhat vertical profile
\crv{solid lines}, as a function of the altitude above the
mid-plane ($z=0$), shown at $r=4\,\kpc$ for various width-scales
$h$. To compare with, the \crv{dotted line} shows the gradient
behavior in the limit of the infinitesimally thin disk model
($h\to0$). The convergence to that limit is point-wise continuous,
although not uniform -- the gradient for $h=0$ is discontinuous at
$z=0$. For higher altitudes, beyond the main concentration of
masses, the gradient behavior is universal.} }\end{figure}

\figref{fig:grad3} shows the behavior of the gradient as a
function of $z$ at $r = 4\,\kpc$ for the \mexhat profile with
various width-scales $h$. The smaller is $h$ the higher is the
gradient's absolute value, but already for $\abs{z}>0.4\,\kpc$ the
gradient is almost independent of $h$.
 For all
finite-width disk, the smaller is the width-scale the \refrm{lower
is the altitude above the mid-plane} at which the high gradient
values characteristic of the infinitesimally disk are obtained.
With increasing and large enough $z$, the gradient in finite-width
disks gradually overlaps with that of the infinitesimally thin
disk. For sufficiently large $\abs{z}$, the differences between
the predictions for disks with various $h$ cease to be visible
\new{and a universal asymptotics can be seen. (Note, that the
total disk mass depends on $h$, which explains tiny differences in
the asymptotics, which could be formally eliminated by rescaling
masses of all disks to the same value.) Physically, this behavior
is clear: for altitudes high enough, beyond the main concentration
of mass in the mid-plane vicinity, the gravitational field of the
infinitesimally thin disk of comparable mass, perfectly
approximates that of a finite-width disk.}
\new{Mathematically, this behavior becomes clear by examining the
asymptotics of the integral \eqref{eqn:gradient_thindisk}, as
shown in the Appendix.}

\new{
In comprehending the behavior of the gradient lines, it is useful
to remember, that the infinitesimally thin disk can be considered
as a limit $h\to0$ of finite-width disks of various vertical
profiles. This property is readily seen in figure
\figref{fig:grad3}. The gradient lines of finite-width disks
converge point-wise to the gradient lines of the infinitesimally
thin disk, although this convergence is not uniform. Owing to this
fact, the gradient lines of finite-width disks are globally
continuous, but there is a discontinuity at $z=0$ in the gradient
lines for the infinitesimally thin disk. From this convergence and
the universal asymptotics referred to above, one can also infer
the presence of the turn overs of the gradient lines for
finite-width disks (the reasoning is given in the Appendix).}

\figref{fig:grad2} shows the gradient as a function of radius at
fixed $z$ for the \mexhat profile with various width-scales $h$,
and additionally, for the exponential vertical profile ($h' =
88\pc$). Similarly as before, the thinner is a disk, the higher
are the the gradient absolute values. At the same time the
difference between the \mexhat vertical profile and the
exponential one is negligible. \newII{From this figure one can
also see that the radial variations of the vertical gradients
reflect the behavior of the radial gradients of the rotation
curve.}
\begin{figure} \centering
\includegraphics[trim = 0mm 14mm 0mm
0mm, clip,width=0.5\textwidth]{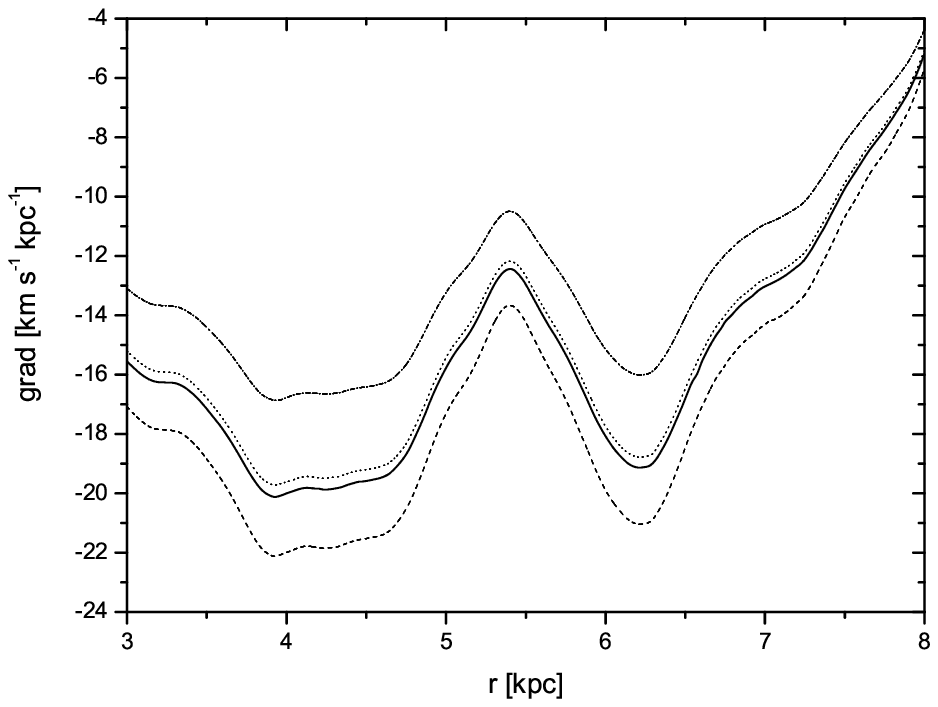}
\includegraphics[trim = 0mm 14mm 0mm 4mm,
clip,width=0.5\textwidth]{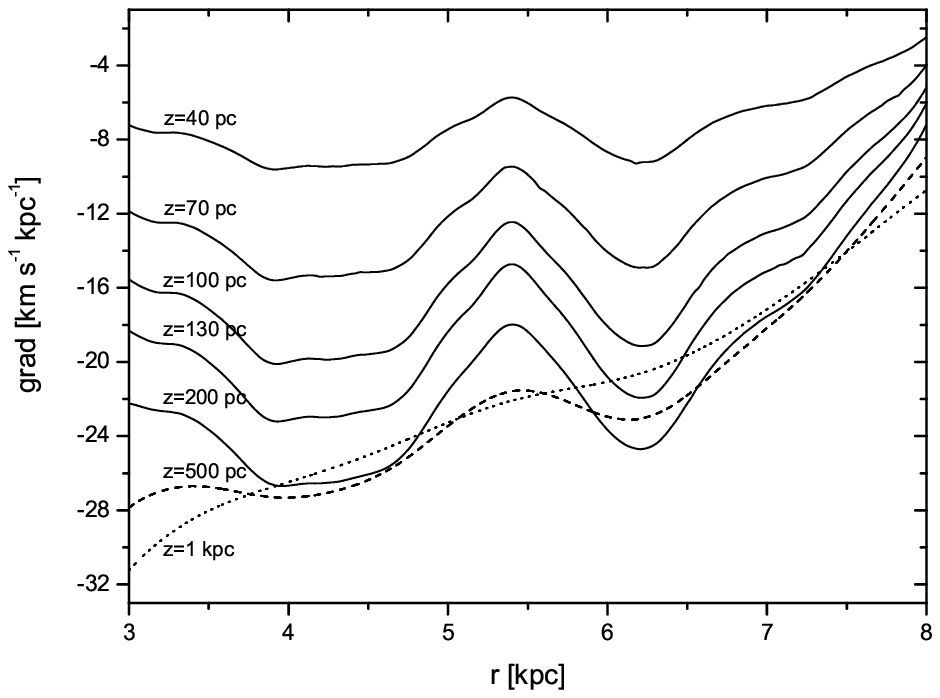}
\includegraphics[trim = 0mm 14mm 3.5mm
6mm, clip,width=0.482\textwidth]{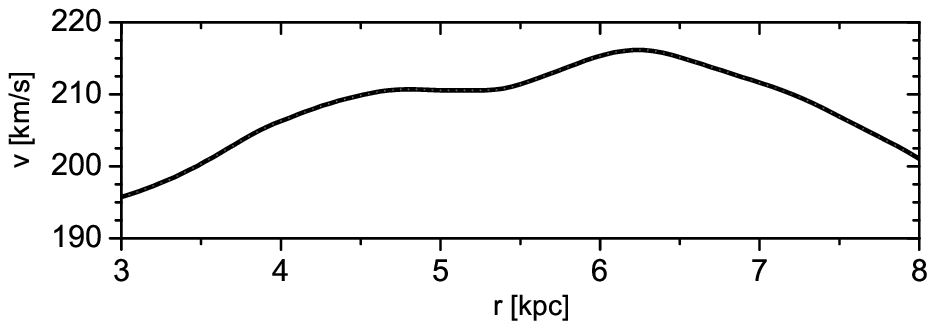}
\includegraphics[trim = 2mm 6mm -17.0mm
6mm, clip,width=0.58\textwidth]{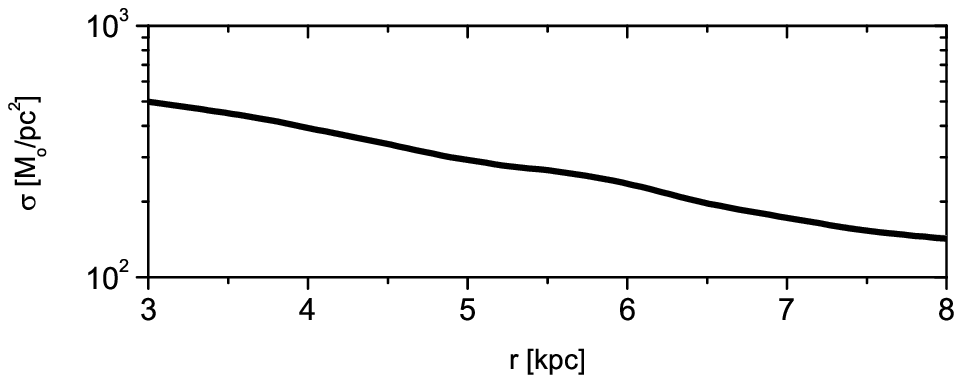}
\caption{\label{fig:grad2} The vertical gradient of rotation of
\mw in a finite-width disk model with \mexhat vertical profile as
a function of the radial variable, shown \refrm{for various
width-scales $h$ and altitudes $z$ above the mid-plane.
\textbf{[Top panel]}: the gradient at an altitude of $100\pc$};
\crv{dashed line} -- $h = 100\pc$, \crv{solid line} -- $h =
117\pc$,  and the uppermost \crv{dash-dotted line} -- $h =
150\pc$. (To compare with, the \crv{dash-dotted line} in the
middle shows the gradient in a disk model with the vertical
exponential profile and the width-scale $h' = 88\pc$ at the same
altitude.)
\new{\textbf{[Middle Top panel]}: the vertical gradient of rotation at
various altitudes for $h=117\pc$.}
\newII{To enable the comparison of structures in the vertical gradient with
those in the rotation curve and in the column mass density,
\textbf{[Middle Bottom panel]} shows the \mw rotation curve, and
\textbf{[Bottom panel]} shows the column mass density for a
finite-width disk model with \mexhat vertical profile ($h =
117\pc$).} }
\end{figure}

\subsection{Comparison of the vertical gradient predictions
with the observations}

The measurements of the vertical gradient in the rotation of our
Galaxy within the radial distance $3-8\,\kpc$ and for altitudes
above the mid-plane out to $100\pc$, give high gradient absolute
values of $22\pm 6\jgrad$ \citep{2008ApJ...679.1288L}. For the
rotation curve used, such high gradients may still be obtained in
more realistic disk models with finite thickness, provided that
the width-scales of the disks are sufficiently small.
\figref{fig:grad3} shows that with the width-scale of $h=117\pc$
for the \mexhat vertical profile the gradient absolute value may
exceed $20\jgrad$. This width-scale is favored by the analysis of
the gravitational microlensing of \secref{sec:microlensing} for
the same rotation curve, assuming $\vsun=200\,\kms$ for the
circular speed of the standard of rest at the Sun position. If
this rotation curve, \new{in conjunction with the finite-width
disk model}, correctly describe the Galactic dynamics, then the
Galactic disk has a small thickness, as suggested by our analysis.
However, an increase by $10\%$ in the rotation velocity due to the
uncertain value of the speed parameter $\vsun$, might result in a
similar increase in the absolute gradient value,
\new{which follows from the scaling of velocities by a (possibly
$r$-dependent) factor}.

It should be noted that some authors obtain high $\vsun$, e.g.,
$239\,\kms$ \citep{2009ApJ...704.1704B} or even higher, as
suggested by other studies. Furthermore, a change in $\vsun$ may
also result in a change in the predicted width-scales of the disk
models: for a higher $\vsun$ we expect a corresponding increase to
occur in the width-scale. \new{E.g., by rescaling the gradients
lines in \figref{fig:grad3} by a factor $>1$, a fixed gradient
value at a fixed altitude will be attained on a gradient line
corresponding to a higher $h$}. Similarly, if the absolute
gradient value reaches the higher end of the range from $16\jgrad$
to $28\jgrad$, allowable by the \citet{2008ApJ...679.1288L}
measurements, the predicted width-scales should become accordingly
larger.

\newIII{
\section{\label{app:toymodel}Column mass density at the Sun vicinity}

\subsection{The rotation curve in axial symmetry}
We can regard the fragment of rotation curve
(\figref{fig:dodatkowy1}) inside solar radius as reliable. It is
determined using a tangent point method applied to rotation data
with relatively small scattering. Most importantly, no mass model
was involved. Under the assumption of axial symmetry (concentric
circular orbits), the tangent point method locates osculating
points at extrema in the Doppler image along lines of constant
galactic longitude (the method distinguishes the observer at the
Sun position).  The uncertainty in the resulting rotation curve
lies mainly in the free parameters $\rsun$ and $\vsun$ which must
be taken from elsewhere. For consistency with the assumptions of
the method, the resulting rotation curve must be modelled under
axial symmetry (non-axisymmetric features in the rotation or in
the derived quantities, such as the predicted column mass density,
are outside the scope of this method).

Contrary to the situation inside solar circle, the rotation curve
outside solar circle is not reliable, the rotation measurement
data are characterized by large scattering and are interpreted
within an assumed mass model on which the resulting external
rotation curve is largely dependent.

\subsection{Comparison of the local measurement
with a prediction of axisymmetric model}

The locally measured column mass density from star counts in the
solar vicinity (a calm an empty region) is unlikely to be
representative for the entire solar circle. The real column
density is not axisymmetric, mostly due to the spiral structure.
This factor should be taken into account when comparing a local
measurement with a model prediction. In particular, the locally
measured value of the column mass density does not have to agree
with that predicted at the Sun position in the framework of
axisymmetric disk model. This only partly explains the discrepancy
between the locally determined value at the Sun vicinity and that
inferred from the rotation curve in the axisymmetric disk model.

The locally determined value of $\approx71\pm\,\msun\pc^{-2}$ (all
gravitating matter below $|z|<1.1\,\kpc$) was inferred from
solving the vertical
 Jeans equation for a stellar
tracer population \citep{1991ApJ...367L...9K}. In a recent paper
\citep{2013ApJ...772..108Z} a similar analysis implies that the
total gravitating column mass density is
$\approx67\,\msun\pc^{-2}$ ($|z|<1.0\,\kpc$), of which the
contribution from all stars is $\approx 42\,\msun\pc^{-2}$ and
that from cold gas is $13\,\msun\pc^{-2}$. The value of
$\approx140\,\msun\pc^{-2}$ inferred in disk model in this paper
should be compared with the local value
$\approx70\,\msun\pc^{-2}$, since the disk model describes total
dynamical mass accounting for the rotation curve.

Another contribution to this discrepancy, which seems more
important, may point to problems with the rotation curve outside
the solar circle, as we illustrate below with the help of a toy
model.

\textbf{A toy model.} For a flattened mass distribution, the
relation between the rotation curve and the resulting column mass
density is nonlocal. The local density value is strongly dependent
on the local radial gradient in the rotation curve, and it also
depends on the external part of rotation curve. On larger scales
this relation is less important and the amount of mass counts
more, like in a spherical model. In consequence of this, the local
value of the surface density at the solar circle will depend on
the accuracy of determining the rotation curve outside solar
circle.

This influence is illustrated in \figref{fig:model}. To obtain the
part of the toy model rotation curve outside solar circle, we
moving-averaged the scattered data $(r, \sqrt{\sum_i^N\frac{r_i
{v_c^2}_i}{N\,r}})$ for $20\,\kpc>r>7\,\kpc$ (open circles in
\figref{fig:model}), taking the $r>8\,\kpc$ part only. Inside the
solar circle, the rotation curve was left unchanged as it is well
determined. We joined the internal and external part so as to
satisfy the constraint $200\,\km/sec$ at $8\,\kpc$. For such
obtained rotation curve, the density at the solar position
$r=8\,\kpc$ reduces to a value $\approx70\,\msun\pc^{-2}$.

We see from this example, that the problem of discrepancy in the
density lies mainly in the rotation curve close to and outside the
solar circle. It follows, that in preparing a better rotation
curve in the future, it may be necessary taking into account
constraints from the local mass density. Clearly, with the
original rotation curve the density is greater. For a less
flattened curve, the local density could be made to agree with the
low value of about $70\,\msun\pc^{-2}$ at $8\,\kpc$, consistently
with results of \citep{1991ApJ...367L...9K} and
\citep{2013ApJ...772..108Z}.
\begin{figure}[h]
\centering
\includegraphics[trim = 6mm 6mm 6mm 6mm,clip,width=0.4\textwidth]{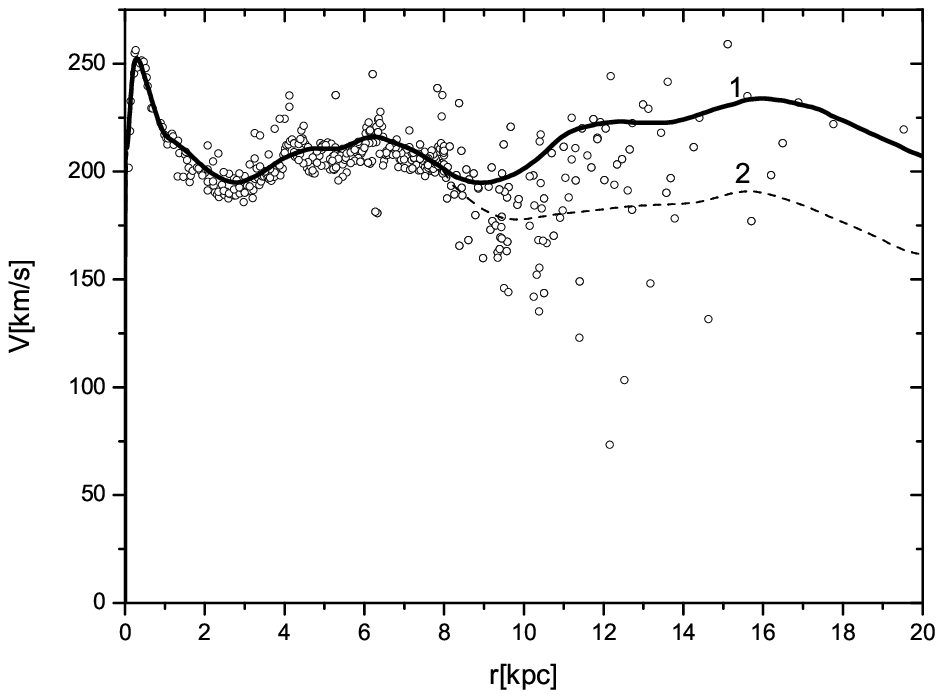}\\
\includegraphics[trim = 7.0mm 6mm 6mm 5mm, clip,width=0.4\textwidth]{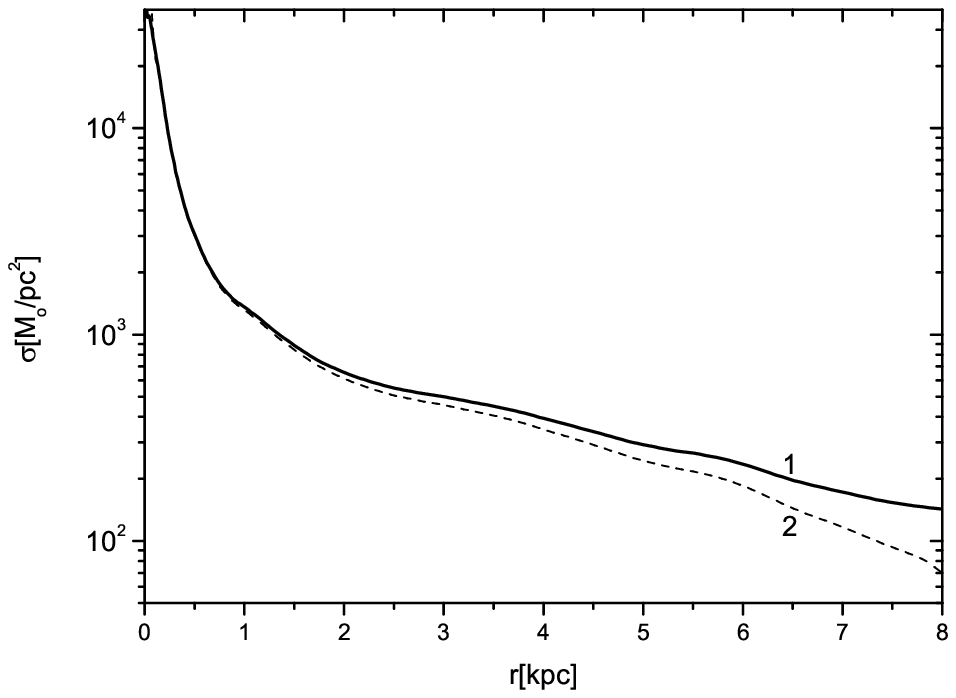}
\caption{\label{fig:model} \newIII{Comparison of column mass
densities for two rotation curves which are identical inside solar
circle and differ from each other outside the circle: \textbf{[Top
panel]} -- 1\textit{[solid line]}: the original rotation curve as
in Fig.\ref{fig:dodatkowy1}, 2\textit{[dashed line]}: the toy
model rotation curve described in the text; \textbf{[Bottom
panel]} -- the corresponding column mass densities for a
finite-width disk with the \mexhat vertical profile and height
scale of $117\pc$.}}
\end{figure}
With the toy model curve, the results for the vertical gradient of
rotation inside solar circle is not changed significantly in a
large region, as shown in Fig.\ref{fig:grad}, even though the
predicted column density at the solar circle is significantly
changed, by a factor of $2$.
\begin{figure}[h]
\centering
\includegraphics[trim = 6mm 6mm 6mm 6mm, clip,width=0.4\textwidth]{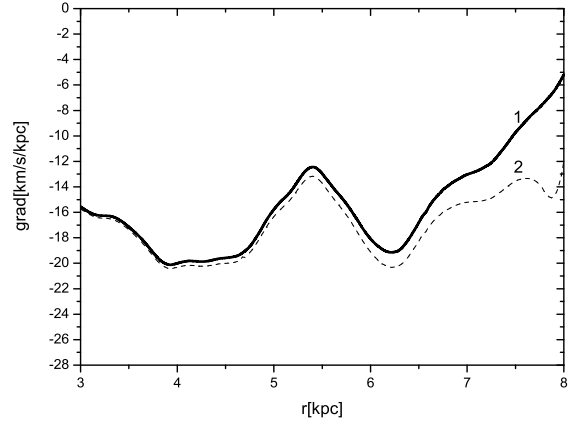}
\caption{\label{fig:grad} \newIII{The vertical gradient of
rotation at $z=100\,\kpc$ for the original rotation curve and for
the toy model curve. The gradient prediction is still high and not
significantly changed in a region $0-7\,\kpc$, compared to its
absolute value. (The local minimum close to $8\,\kpc$ is a result
of a discontinuity in the toy model curve).}}
\end{figure}
This change would not influence much the results well inside the
solar circle also for quantities which are integrals over a large
region, such as the integrated optical depth we modelled in the
present paper. }

\section{Summary and concluding remarks}

\add{

We obtained mass distribution in finite-width disks in an
iterative fashion from a given rotation curve. We assumed that the
volume density $\rho(r,z)$ could be factorized as
$\rho(r,z)=\sigma(r)f(z)$, where $f(z)$ is a normalized vertical
profile, with a characteristic width-scale which is a free
parameter. We did not assume any constraints on the column mass
density $\sigma(r)$, so that its functional dependence was
governed entirely by the shape of the rotation curve.
\new{The density effectively describes all forms of the dynamical mass
inferred from the Galactic rotation, therefore, the disk-width
should not be confused with the width-scale parameters measured in
the Sun vicinity for various stellar subsystems.}
\footnote{\new{Incidentally, the obtained width-scale of
$\approx100\pc$ is consistent with the distribution of
\newII{massive stars} in the Sun vicinity. \newII{However,
massive} \newII{stars are
 rare locally and do not contribute substantially to the mass for any
 proposed IMF.}}}

\new{Based on
the gravitational microlensing measurements, it was possible
within the disk model framework to determine the effective
width-scales. We tested such obtained mass distributions by
comparing the model predictions for the vertical gradient of the
azimuthal velocity with the gradient measurements.}

The (integrated) optical depth from the microlensing measurements
is influenced by the amount of mass distributed along the lines of
sight towards the Galactic center, whereas the details of the
distribution are less important. We inferred the optimum
width-scales of the considered disks by means of finding the best
fits to the optical depth measured along various lines of sight.
This shows that microlensing can be used as a tool to
independently constrain the mass distribution models. With such
determined width-scales, the resulting prediction for the behavior
of the vertical gradient of rotation was compared with the
gradient measurements in the mid-plane vicinity. This comparison
turned out consistent with a small disk thickness.
\new{(Interestingly enough, for such determined width-scales, the
effective disk widths defined by the \Q{1/e criterion}, are almost
equal: $2h'=176\pc$ and $\sim1.49h=174\pc$, respectively, for the
exponential and for the \mexhat vertical profiles.)}

The behavior of the vertical gradient of the azimuthal velocity
and its value, when calculated on low altitudes above the
mid-plane, is very sensitive to the width-scale parameter. At a
given altitude in the gradient measurements region, the calculated
gradient value changes significantly with the width-scale
parameter. When the parameter is too high the absolute gradient
value is too low compared with the measurements.   Higher absolute
gradients at low altitudes above the mid-plane suggest a smaller
effective thickness of the Galactic disk. The rotation velocity is
another factor that governs the gradient value. In particular,
given a disk thickness and the gradient behavior, one could
constrain the allowable range for the motion of the standard of
rest at the Sun position (the width-scale could be increased for
higher $\vsun$). Testing the vertical structure of the mass
distribution with the help of the gradient measurements is thus a
particularly sensitive tool, and this is therefore important to
have high accuracy measurements of the gradient at small altitudes
above the mid-plane. }

Column mass density of flattened mass distributions is sensitive
to uncertainties in the circular velocity. This sensitivity can be
observed in the approximation of infinitely thin disk model
\citep{1987gady.book.....B}. It is therefore important to have
reliable rotation curves when studying flattened galaxies.
However, Galactic rotation is relatively well known inside solar
circle, therefore, with better data, we can expect some
differences in the column density to occur close to or outside
solar circle. These changes should not influence significantly
results for global quantities inside solar circle, such as the
predicted integrated optical depth \newIII{or vertical gradient
inside the solar circle. In particular, as shown in
\secref{app:toymodel}, with suitably changed external part of the
rotation curve,  the disk model prediction for the column density
at the solar circle of $\approx140\,\msun\pc^{-2}$ made with the
present rotation curve, could be reduced to a local value of
$\approx70\,\msun\pc^{-2}$ inferred at the Sun vicinity from Jeans
modelling \citep{1991ApJ...367L...9K,2013ApJ...772..108Z}.  }

\section*{Acknowledgements}
 We would like to thank the anonymous referee for
a careful reading of our manuscript and for detailed and
constructive suggestions that improved the presentation of our
results.

\begin{appendix}
\section{}
\subsection{Derivation
of \eqref{eqn:rotation_thindisk} and \eqref{eqn:gradient_thindisk}
}

\begin{figure}
\centering
\includegraphics[trim= 0mm 0mm 0mm 0mm, clip,width=0.5\textwidth]{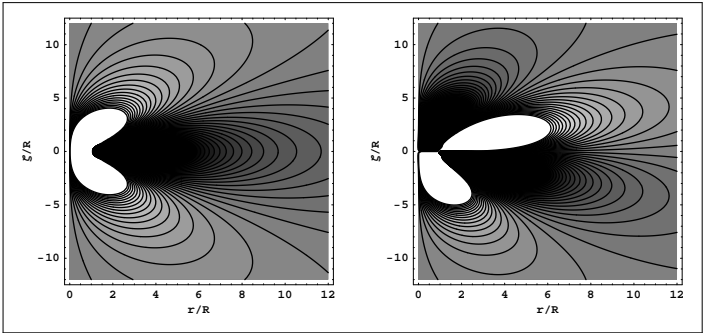}
\caption{\label{fig:kernel} Contour plots of
$\tanh\br{J(r,R,\zeta)}$ and $\tanh\br{
R\,\partial_{\zeta}J(r,R,\zeta)}$. These functions are singular at
$r=R$ and $\zeta=0$ and continuous elsewhere. }
\end{figure}

The kernel function $J$ defined in the introduction is singular at
an isolated point $\br{\zeta=0, r=R}$, and is continuous
elsewhere. This singularity is integrable in the principal value
sense in \eqref{eqn:rotation_thindisk} and
\eqref{eqn:gradient_thindisk}. Function $J$ is scale-invariant. In
particular, $J(r,R,\zeta)=J(r/R,1,\zeta/R)$, which means that $J$
is effectively dependent on two variables only: $r/R$ and
$\zeta/R$. This property allows us to represent $J$ on a plane as
in \figref{fig:kernel}.

Differentiation and taking limits can be interchanged with the
integration only under particular conditions imposed on the
function under the integration sign. If not said otherwise, we
assume these conditions are met. For this reason, the case of
infinitesimally thin disk (with $f(z)=\delta(z)$) must be treated
separately.

The expression for $R\,\partial_R\Psi(R,Z)$ in
\secref{sec:DiskTheory} involves an integral
$$\mathcal{I}=\int\limits_{-\infty}^{\infty}\!\!\ud{z}\,
\rho(r,z)J(r,R,z-Z).$$ The integral can be written as a sum
$\mathcal{I}=\int\limits_{-\infty}^{0}\!\!\ud{z}\,
\rho(r,z)J(r,R,z-Z)+\int\limits_{0}^{\infty}\!\!\ud{z}\,
\rho(r,z)J(r,R,z-Z)$ if the two summands exist. By substituting
$z\to-\tilde{z}$, the first integral can be rewritten as
$\int\limits_{+\infty}^{0}\!\!\br{-\ud{\tilde{z}}}\,
\rho(r,-\tilde{z})J(r,R,-[\tilde{z}+Z])$, and next, since
$\rho(r,-\zeta)=\rho(r,\zeta)$ and $J(r,R,-\zeta)=J(r,R,\zeta)$,
as $\int\limits_{0}^{+\infty}\!\!\ud{\tilde{z}}\,
\rho(r,\tilde{z})J(r,R,\tilde{z}+Z)$. Finally, by renaming
$\tilde{z}\to z$, we obtain
$$\mathcal{I}=\int\limits_{0}^{\infty}\!\!\ud{z}\,
\rho(r,z)\cdot\sq{J(r,R,z+Z)+J(r,R,z-Z)}.$$ This proves
\eqref{eqn:rotation_thindisk}.

\medskip

The integral expression for the vertical
gradient of rotation in the quasi-circular
orbits approximation can be proved by
performing a partial differentiation of
$\mathcal{I}$ under the integration sign:
$$\partial_Z\mathcal{I}=\int\limits_{0}^{\infty}\!\!\ud{z}\,
\rho(r,z)\cdot\sq{\partial_ZJ(r,R,z+Z)+\partial_ZJ(r,R,z-Z)}.$$
Now, $\partial_ZJ(r,R,z\pm
Z)=\pm\partial_zJ(r,R,z\pm Z)$, which
implies that
$\partial_Z\mathcal{I}=\int\limits_{0}^{\infty}\!\!\ud{z}\,
\rho(r,z)\cdot\sq{\partial_zJ(r,R,z+Z)-\partial_zJ(r,R,z-Z)}$.
Integration by parts under usual conditions
gives
$\partial_Z\mathcal{I}=\mathcal{B}-\int\limits_{0}^{\infty}\!\!\ud{z}\,
\partial_z\rho(r,z)\cdot\sq{J(r,R,z+Z)-J(r,R,z-Z)}$,
where $\mathcal{B}=\br{
\lim\limits_{z\to+\infty}\rho(r,z)\sq{J(r,R,z+Z)-J(r,R,z-Z)}}-
\br{\rho(r,0^+)\sq{J(r,R,Z)-J(r,R,-Z)}}$ is a boundary term. The
expression in the first round bracket in this term is zero for any
finite $Z,R,r$, on account of the vanishing of the difference in
the square bracket and the vanishing of $\rho$ in the same limit
(it suffices to assume a finite support of $\rho$). The expression
in the second round bracket also vanishes for finite $\rho(r,0)$
since $J$ is an even function of the third argument. Hence, the
boundary term $\mathcal{B}$ also vanishes. Formally, one should
also verify whether the integration with respect to $r$ and taking
the above limit $z\to+\infty$ commute, as there is an integration
over $r$ present in the expression for $R\,\partial_R\Psi$. When
this requirement is met, we have
$$\partial_Z\!\!\int\limits_{0}^{\infty}\ud{r}\,\mathcal{I}=
\int\limits_{0}^{\infty}\!\!\ud{r}\!\!
\int\limits_{0}^{\infty}\!\!\ud{z}\,\,
\partial_z\rho(r,z)\cdot\sq{J(r,R,z-Z)-J(r,R,z+Z)},$$
which proves \eqref{eqn:gradient_thindisk}.

\subsection{A special case: the exponential vertical profile}

In the case of the vertical
exponential falloff of the density
profile,
$\rho(r,z)=\frac{\exp{-|z|/h}}{2h}\,\sigma(r)$,
calculation of integrals
\eqref{eqn:rotation_thindisk} and
\eqref{eqn:gradient_thindisk} can be
simplified. Then, $
v^2_{\phi}(R,Z)=G\!
\int\limits_{0}^{\infty}\!\!\ud{r}\!\!\int\limits_{0}^{\infty}\!\!\ud{z}\,
\sigma(r)\,\frac{\exp{-z/h}}{2h}\cdot\sq{J(r,R,z-Z)+J(r,R,z+Z)}$
and
$\partial_Zv_{\phi}(R,Z)=\frac{G}{v_{\phi}(R,Z)}
\int\limits_{0}^{\infty}\!\!\ud{r}\!\!\int\limits_{0}^{\infty}\!\ud{z}
\,\sigma(r)\,\frac{-\exp{-z/h}}{2h^2}\cdot\sq{J(r,R,z-Z)-J(r,R,z+Z)}$.
Now, both $v_{\phi}(R,Z)$ and
$\partial_Zv_{\phi}(R,Z)$ can be
expressed in terms of two integrals
$I_+$ and $I_-$:
$$I_{\pm}=\frac{G}{h}
\int\limits_{0}^{\infty}\!\!\sigma(r)\,\ud{r}\!\!\int\limits_{0}^{\infty}\!\ud{z}
\,\exp{-z/h}\,J(r,R,z\pm Z).$$ Namely, for
$\rho(r,z)=\frac{\exp{-|z|/h}}{2h}\,\sigma(r)$,
$v_{\phi}(R,Z)=\sqrt{I_++I_-}$ and $\partial_Zv_{\phi}(R,Z)=
\frac{I_+-I_-}{2\,h\sqrt{I_++I_-}}$.

\subsection{Qualitative properties of the
vertical gradient of rotation and the presence of \Q{turn-overs}.}

\eqref{eqn:gradient_thindisk} involves an integral
$\int\limits_{0}^{\infty}\ud{z}f'(z)\sq{J(r,R,z-Z)-J(r,R,z+Z)}$
which equals
$\mathcal{K}=\int\limits_{-\infty}^{\infty}\ud{z}f'(z)J(r,R,Z-z)$,
owing to the symmetry of $J$ and $f$. Now, consider $|Z|>A>0$
large enough, beyond the main mass concentration, such that
$\int\limits_{-A}^{A}\!\!f(z)\,\ud{z}\approx 1$ (by assumption
$\int\limits_{-\infty}^{+\infty}\!\!f(z)\,\ud{z}= 1$). For such
$Z$, only a region $|z|\ll|Z|$ contributes to $\mathcal{K}$, and
we can use  the approximation formula
$J(r,R,Z-z)-J(r,R,Z)\approx-z\,\partial_{Z}J(r,R,Z)$ to obtain
$\mathcal{K}\approx
J(r,R,Z)\int\limits_{-\infty}^{+\infty}f'(z)\ud{z}-
\partial_ZJ(r,R,Z)\int\limits_{-\infty}^{+\infty}zf'(z)\ud{z}$.
Since $f(z)$ and $z f(z)$ vanish at the infinity, integrating by
parts gives $\mathcal{K}\approx
\partial_ZJ(r,R,Z)\int\limits_{-\infty}^{+\infty}f(z)\ud{z}=\partial_ZJ(r,R,Z)$.
On the other hand, for $f(z)=\delta(z)$ and for $Z\ne0$,
$\mathcal{K}=\int\limits_{-\infty}^{\infty}\ud{z}\delta'(z)J(r,R,Z-z)=
-\int\limits_{-\infty}^{\infty}\ud{z}\delta(z)\partial_zJ(r,R,Z-z)=
\partial_ZJ(r,R,Z)$.
Hence, we obtain an intuitively clear result: {\textit{for all
finite-width thin disks with the same column mass density
$\sigma(r)$, the behavior of the vertical gradient at altitudes
large enough, is universal and the same as for an infinitesimally
thin disk with surface mass density $\sigma(r)$.}}

 Another
qualitative result is obtained in the limit $Z\to0$. For $Z\ne0$,
$\mathcal{K}(R,r,-Z)=-\mathcal{K}(R,r,Z)$, thus
$\lim\limits_{Z\to0}\mathcal{K}=0$ by continuity of $\mathcal{K}$
as a function of $Z$, and the vertical gradient is zero at $Z=0$,
at least for those mass distributions for which the usual theorems
on the continuity of functions defined by integrals
\eqref{eqn:gradient_thindisk} apply.

However, there is an exception from the above continuity behavior
of the gradient lines at $Z=0$. It is important to remember that
the operation of taking various limits and the operation of
integration are not interchangeable in general. In particular, an
integral of a function sequence consisting of continuous functions
with a parameter can result in a discontinuous function of that
parameter. For $f(z)=\delta_n(z)$, where $\delta_n$ is a
functional sequence representing the Dirac $\delta$, the result of
continuity of the gradient does not necessarily follow and we can
have a nonzero value in the same limit, in which case the integral
\eqref{eqn:gradient_thindisk} is discontinuous at $Z=0$. To give a
simple example of what then may happen, consider a function
sequence
$u_{n}(x,y)=\frac{2xy^2}{\pi(1+x^2)(\frac{1}{n^2}+x^2+y^2)^2}$,
then $\int\limits_{-\infty}^{+\infty} u_n(x,y)\ud{y}=g_n(x)$, with
$g_n(x)=\frac{n x}{(1+x^2)\sqrt{1+n^2 x^2}}$. Now, consider a
limiting function $g(x)=\lim\limits_{n\to+\infty}g_n(x)$ and see
if it is continuous at $x=0$. For $x=0$,
$g(0)\equiv\lim\limits_{n\to+\infty}g_n(0)=0$, whereas for
$x\ne0$,
$g(x)\equiv\lim\limits_{n\to+\infty}g_n(x)=\frac{1}{1+x^2}\ne0$,
thus, $g(0)=0\ne1=\lim\limits_{x\to0} g(x)$, therefore $g(x)$ is
discontinuous at $x=0$. Now, think of the gradient lines in
\figref{fig:grad3} -- then the finite disk corresponds to the
situation described by $g_n(x)$ ($h=n^{-1}>0$), whereas the
infinitesimally thin disk corresponds to the situation of
discontinuous $g(x)$ ($h=0$).

Finally, we may try to understand the occurrence of the
\Q{turn-overs} in the gradient lines for $h>0$, such as those seen
in \figref{fig:grad} or \figref{fig:grad3}. First, note that a
gradient line must asymptotically converge to zero, which is the
universal asymptotics property discussed earlier. Second, the
gradient line starts from $0$ at $Z=0$, which we have also seen
above. Now, let's perform a mapping of the region $0<Z<+\infty$ to
an interval $0<Z<1$ by means of a transformation $Z\to\tanh{Z}$.
Then the transformed gradient lines are continuous for $0\leq
Z\leq 1$ and vanish at the boundaries. Next, we apply the Rolle's
theorem on continuously differentiable functions vanishing on the
boundaries of a compact and simply connected interval, and we
infer that there must be at least one point inside the interval
where the gradient line has a local minimum, which explains the
presence of a \Q{turn-over}. The Rolle's theorem does not apply to
gradient lines of the infinitesimally thin disk,  because of the
discontinuity, and the analogous \Q{turn-overs} do not have to
occur, which is the case in \figref{fig:grad} or
\figref{fig:grad3}.

\end{appendix}

\bibliography{grubydyskgradmikrosocz_v4}

\begin{thebibliography}{}

\bibitem[\protect\citeauthoryear{{Alard}}{{Alard}}{1997}]{1997A&A...321..424A}
{Alard} C.,  1997, \aap, 321, 424

\bibitem[\protect\citeauthoryear{{Alcock}, {Allsman}, {Alves}, {Axelrod} \& {at
  al}}{{Alcock} et~al.}{2000}]{2000ApJ...542..281A}
{Alcock} C.,  {Allsman} R.~A.,  {Alves} D.~R.,  {Axelrod} T.~S.,    {at al}
  2000, \apj, 542, 281

\bibitem[\protect\citeauthoryear{{Binney} \& {Tremaine}}{{Binney} \&
  {Tremaine}}{1987}]{1987gady.book.....B}
{Binney} J.,  {Tremaine} S.,  1987, {Galactic dynamics}

\bibitem[\protect\citeauthoryear{{Bissantz}, {Debattista} \&
  {Gerhard}}{{Bissantz} et~al.}{2004}]{2004ApJ...601L.155B}
{Bissantz} N.,  {Debattista} V.~P.,    {Gerhard} O.,  2004, \apjl, 601, L155

\bibitem[\protect\citeauthoryear{{Bissantz} \& {Gerhard}}{{Bissantz} \&
  {Gerhard}}{2002}]{2002MNRAS.330..591B}
{Bissantz} N.,  {Gerhard} O.,  2002, \mnras, 330, 591

\bibitem[\protect\citeauthoryear{{Bovy}, {Hogg} \& {Rix}}{{Bovy}
  et~al.}{2009}]{2009ApJ...704.1704B}
{Bovy} J.,  {Hogg} D.~W.,    {Rix} H.-W.,  2009, \apj, 704, 1704

\bibitem[\protect\citeauthoryear{{Bratek}, {Ja{\l}ocha} \&
  {Kutschera}}{{Bratek} et~al.}{2008}]{2008MNRAS.391.1373B}
{Bratek} {\L}.,  {Ja{\l}ocha} J.,    {Kutschera} M.,  2008, \mnras, 391, 1373

\bibitem[\protect\citeauthoryear{{Derue}, {Afonso}, {Alard}, {Albert},
  {Andersen}, {Ansari}, {Aubourg}, {Bareyre}, {Bauer}, {Beaulieu}, {Blanc},
  {Bouquet} \& {at al}}{{Derue} et~al.}{2001}]{2001A&A...373..126D}
{Derue} F.,  {Afonso} C.,  {Alard} C.,  {Albert} J.-N.,  {Andersen} J.,
  {Ansari} R.,  {Aubourg} {\'E}.,  {Bareyre} P.,  {Bauer} F.,  {Beaulieu}
  J.-P.,  {Blanc} G.,  {Bouquet} A.,    {at al} 2001, \aap, 373, 126

\bibitem[\protect\citeauthoryear{{Gradshtein}, {Ryzhik}, {Jeffrey} \&
  {Zwillinger}}{{Gradshtein} et~al.}{2007}]{Ryzhik}
{Gradshtein} I.,  {Ryzhik} I.,  {Jeffrey} A.,    {Zwillinger} D.,  2007, Table
  of integrals, series and products.
Academic Press

\bibitem[\protect\citeauthoryear{{Hamadache}, {} \& {et al.}}{{Hamadache}
  et~al.}{2006}]{2006A&A...454..185H}
{Hamadache} C.,  {}   {et al.} 2006, \aap, 454, 185

\bibitem[\protect\citeauthoryear{{Ja{\l}ocha}, {Bratek} \&
  {Kutschera}}{{Ja{\l}ocha} et~al.}{2008}]{2008ApJ...679..373J}
{Ja{\l}ocha} J.,  {Bratek} {\L}.,    {Kutschera} M.,  2008, \apj, 679, 373

\bibitem[\protect\citeauthoryear{{Ja{\l}ocha}, {Bratek}, {Kutschera} \&
  {Skindzier}}{{Ja{\l}ocha} et~al.}{2010}]{2010MNRAS.407.1689J}
{Ja{\l}ocha} J.,  {Bratek} {\L}.,  {Kutschera} M.,    {Skindzier} P.,  2010,
  \mnras, 407, 1689

\bibitem[\protect\citeauthoryear{{Ja{\l}ocha}, {Bratek}, {Kutschera} \&
  {Skindzier}}{{Ja{\l}ocha} et~al.}{2011}]{2011MNRAS.412..331J}
{Ja{\l}ocha} J.,  {Bratek} {\L}.,  {Kutschera} M.,    {Skindzier} P.,  2011,
  \mnras, 412, 331

\bibitem[\protect\citeauthoryear{{Kuijken} \& {Gilmore}}{{Kuijken} \&
  {Gilmore}}{1991}]{1991ApJ...367L...9K}
{Kuijken} K.,  {Gilmore} G.,  1991, \apjl, 367, L9

\bibitem[\protect\citeauthoryear{{Levine}, {Heiles} \& {Blitz}}{{Levine}
  et~al.}{2008}]{2008ApJ...679.1288L}
{Levine} E.~S.,  {Heiles} C.,    {Blitz} L.,  2008, \apj, 679, 1288

\bibitem[\protect\citeauthoryear{{Moniez}}{{Moniez}}{2010}]{2010GReGr..42.2047M}
{Moniez} M.,  2010, General Relativity and Gravitation, 42, 2047

\bibitem[\protect\citeauthoryear{{Paczynski}}{{Paczynski}}{1986}]{1986ApJ...304....1P}
{Paczynski} B.,  1986, \apj, 304, 1

\bibitem[\protect\citeauthoryear{{Popowski}, {at al} \& {MACHO
  Collaboration}}{{Popowski} et~al.}{2005}]{2005ApJ...631..879P}
{Popowski} P.,  {at al}   {MACHO Collaboration} 2005, \apj, 631, 879

\bibitem[\protect\citeauthoryear{{Sikora}, {Bratek}, {Ja{\l}ocha} \&
  {Kutschera}}{{Sikora} et~al.}{2012}]{2012A&A...546A.126S}
{Sikora} S.,  {Bratek} {\L}.,  {Ja{\l}ocha} J.,    {Kutschera} M.,  2012, \aap,
  546, A126

\bibitem[\protect\citeauthoryear{{Smith}, {Wo{\'z}niak}, {Mao} \&
  {Sumi}}{{Smith} et~al.}{2007}]{2007MNRAS.380..805S}
{Smith} M.~C.,  {Wo{\'z}niak} P.,  {Mao} S.,    {Sumi} T.,  2007, \mnras, 380,
  805

\bibitem[\protect\citeauthoryear{{Sofue}, {Tutui}, {Honma}, {Tomita},
  {Takamiya}, {Koda} \& {Takeda}}{{Sofue} et~al.}{1999}]{1999ApJ...523..136S}
{Sofue} Y.,  {Tutui} Y.,  {Honma} M.,  {Tomita} A.,  {Takamiya} T.,  {Koda} J.,
     {Takeda} Y.,  1999, \apj, 523, 136

\bibitem[\protect\citeauthoryear{{Spitzer}
  Jr.}{{Spitzer}}{1942}]{1942ApJ....95..329S}
{Spitzer} Jr. L.,  1942, \apj, 95, 329

\bibitem[\protect\citeauthoryear{{Sumi} \& {at al}}{{Sumi} \& {at
  al}}{2003}]{2003ApJ...591..204S}
{Sumi} T.,  {at al} 2003, \apj, 591, 204

\bibitem[\protect\citeauthoryear{{Sumi}, {Wo{\'z}niak}, {Udalski},
  {Szyma{\'n}ski}, {Kubiak}, {Pietrzy{\'n}ski}, {Soszy{\'n}ski},
  {{\.Z}ebru{\'n}}, {Szewczyk}, {Wyrzykowski} \& {Paczy{\'n}ski}}{{Sumi}
  et~al.}{2006}]{2006ApJ...636..240S}
{Sumi} T.,  {Wo{\'z}niak} P.~R.,  {Udalski} A.,  {Szyma{\'n}ski} M.,  {Kubiak}
  M.,  {Pietrzy{\'n}ski} G.,  {Soszy{\'n}ski} I.,  {{\.Z}ebru{\'n}} K.,
  {Szewczyk} O.,  {Wyrzykowski} {\L}.,    {Paczy{\'n}ski} B.,  2006, \apj, 636,
  240

\bibitem[\protect\citeauthoryear{{Wyrzykowski}, {Koz{\l}owski}, {Skowron},
  {Udalski}, {Szyma{\'n}ski}, {Kubiak}, {Pietrzy{\'n}ski}, {Soszy{\'n}ski},
  {Szewczyk}, {Ulaczyk} \& {Poleski}}{{Wyrzykowski}
  et~al.}{2011}]{2011MNRAS.413..493W}
{Wyrzykowski} {\L}.,  {Koz{\l}owski} S.,  {Skowron} J.,  {Udalski} A.,
  {Szyma{\'n}ski} M.~K.,  {Kubiak} M.,  {Pietrzy{\'n}ski} G.,  {Soszy{\'n}ski}
  I.,  {Szewczyk} O.,  {Ulaczyk} K.,    {Poleski} R.,  2011, \mnras, 413, 493

\bibitem[\protect\citeauthoryear{{Zhang}, {Rix}, {van de Ven}, {Bovy}, {Liu} \&
  {Zhao}}{{Zhang} et~al.}{2013}]{2013ApJ...772..108Z}
{Zhang} L.,  {Rix} H.-W.,  {van de Ven} G.,  {Bovy} J.,  {Liu} C.,    {Zhao}
  G.,  2013, \apj, 772, 108

\end{thebibliography}
\bibliographystyle{mn2e}

\end{document}